\documentclass[iop]{emulateapj}

\usepackage{graphicx, natbib, color, bm, amsmath, epsfig, slashbox,ctable}

\newcommand{\avir}{\ensuremath{\alpha_{\rm{vir}}}}
\newcommand{\mach}{\ensuremath{\mathcal{M}}}
\newcommand{\alfmach}{\ensuremath{\mathcal{M_{\rm{A}}}}}
\newcommand{\msun}{\ensuremath{M_\odot}}

\def\rms{\ensuremath{\rm{rms}}}

\newcommand{\p}[1]{#1^\prime }
\newcommand{\pp}[1]{#1^{\prime \prime} }
\newcommand{\tff}{\ensuremath{t_{\rm{ff}}}}
\newcommand{\kkmin}{\ensuremath{k/k_{\rm{min}}}}

\newcommand{\dc}[1]{}
\def\bvec{\ensuremath{\vec{B}}}
\def\vvec{\ensuremath{\vec{v}}}

\newcommand{\sci}[1]{\ensuremath{\times {10^{#1}}}}
\newcommand{\percc}{\ensuremath{\rm{cm}^{-3}}}

\newcommand{\kms}{\ensuremath{\rm{km}\ \rm{s}^{-1}}}

\def\alf{Alfv\' en}
\def\Alfvenic{Alfv\' enic}
\def\sa{super-Alfv\' enic}

\def\suba{sub-Alfv\' enic}

\def\transa{trans-Alfv\' enic}

\def\cs{\ensuremath{c_{\rm{s}}}}
\def\sfrff{\ensuremath{\rm{SFR_{ff}}}}

\newcommand{\Fourier}[1]{\ensuremath{\tilde{#1}}}

\def\hw{0.49}

\def\xcr{\ensuremath{\rho_{\rm{cr}}}}
\def\rhot{\ensuremath{\rho_{\rm{t}}}}

\def\rrex{\ensuremath{n}} 
\def\urex{\ensuremath{\epsilon}} 
\def\vrex{\ensuremath{m}} 
\def\vbex{\ensuremath{p}} 
\def\brex{\ensuremath{q}} 
\def\betarex{\ensuremath{s}} 
\def\prex{\ensuremath{\zeta}} 
\def\psigmaex{\ensuremath{\lambda}} 
\def\pvex{\ensuremath{\nu}} 

\def\betao{\ensuremath{\beta_0}}
\def\betah{\ensuremath{\beta_0=20}}
\def\betam{\ensuremath{\beta_0=2}}
\def\betal{\ensuremath{\beta_0=0.2}}
\def\tlate{\ensuremath{t=0.6\tff}}
\def\tmid{\ensuremath{t=0.3\tff}}
\def\tearly{\ensuremath{t=0.1\tff}}

\def\tdyn{\ensuremath{t_{\rm{dyn}}}}
\def\rholow{\ensuremath{\rho_{\rm{t}}}}
\def\rhohigh{\ensuremath{\rho_{\rm{r}}}}
\def\betadyn{\ensuremath{\beta_{\rm{dyn}}}}
\def\betath{\ensuremath{\beta_{\rm{th}}}}
\def\rhoc{\ensuremath{\rho_{\rm{cr}}}}
\renewcommand\vvec{\ensuremath{ {\bf v}}}
\renewcommand\bvec{\ensuremath{ {\bf B}}}
\def\vc{\ensuremath{{\bf v_c}}}
\def\vs{\ensuremath{{\bf v_s}}}
\def\betarespectively{$\betal,\ 2$, and $20$, respectively}
\newcommand{\hltwo}[1]{#1}
\newcommand{\hlthree}[1]{#1}

\begin{document}
\title{The Two States of Star-Forming Clouds}
\author{David Collins\altaffilmark{1,2},
Alexei G. Kritsuk\altaffilmark{1}, 
Paolo Padoan\altaffilmark{3}, 
Hui Li\altaffilmark{2},
Hao Xu\altaffilmark{2},
Sergey D. Ustyugov\altaffilmark{4},
Michael L. Norman\altaffilmark{1} 
}

\altaffiltext{1}{Center for Astrophysics \& Space Sciences and
Department of Physics, University of California at San Diego, La Jolla, CA}
\altaffiltext{2}{Theoretical division, Los Alamos National Lab, Los Alamos, NM}
\altaffiltext{3}{ICREA-ICC, University of Barcelona, Spain }
\altaffiltext{4}{Keldysh Institute for Applied Mathematics, Russian Academy of
Sciences, Miusskaya Pl. 4, Moscow 125047, Russia}

\begin{abstract}
We examine the effects of self-gravity and magnetic fields on supersonic
turbulence in isothermal molecular clouds with high resolution simulations and
adaptive mesh refinement.
These simulations use large root grids ($512^3$) to capture turbulence
and four levels of refinement to capture high density, for an
effective resolution of $8,196^3$.  Three Mach 9
simulations are performed, two \sa\ and one \transa.  We find
that gravity splits the clouds into two
populations, one low density turbulent state and one high density collapsing
state.  The low density state exhibits
properties similar to non-self-gravitating in this regime, and we examine the
effects of varied magnetic field strength on statistical properties:  the density probability
distribution function is approximately lognormal; velocity power spectral slopes
decrease with field strength; alignment between velocity and magnetic field
increases with field; the magnetic field
probability distribution can be fit to a stretched exponential.  
The high density state
is characterized by self-similar spheres;
 the density PDF is a power-law;
collapse rate decreases with increasing mean field;  
density power spectra have positive slopes,
$P(\rho,k)\propto k$; thermal-to-magnetic pressure ratios are unity for
all simulations; dynamic-to-magnetic pressure ratios are larger than
unity for all simulations;  
magnetic field distribution is a power-law.  
The high \alf\ Mach numbers in collapsing regions explain recent observations of magnetic influence
decreasing with density.  We also find that the high density state is 
found in filaments formed by converging flows, consistent with recent
\emph{Herschel} observations.  Possible modifications to existing star formation theories
are explored.  
\end{abstract}


\maketitle

\newif\ifpdf
\pdftrue
\section{Introduction}
Star formation is one of the most important outstanding problems in astronomy
and astrophysics.  Over the last six decades, the theory of star formation has
gone through a major paradigm shift.  Early work \citep{Mestel56, Mouschovias76}
focused on the dominance of magnetic fields as the primary physical agent in
star formation, suppressing collapse to support the perceived long lifetime of
molecular clouds.  Later work \citep{Larson81, Elmegreen93, Padoan02,
Krumholz05,Padoan11} shifted the focus from magnetically dominated collapse to
turbulence dominated collapse.  In a swing in the other direction, new
observations \citep{Li09, Crutcher10, Heyer12, Targon11} have indicated that at certain
size and density scales, magnetic fields dominate, but at smaller scales the field importance is
reduced.  

This paradigm shift towards turbulence has been made possible in large part due to
the ever increasing capability of computers and 
magnetohydrodynamic (MHD) algorithms, which allow increasingly accurate
simulations of MHD turbulence \citep{Kritsuk11b}.  Recent algorithm progress has been made by
including magnetic fields in high dynamic range codes, most notably adaptive
mesh refinement (AMR) \citep{Balsara01, Fromang06, Collins11} and smoothed
particle hydrodynamics  \citep{Price04, Dolag09,  Gaburov11, Price12}. 

The three most important physical agents in star formation are gravity,
turbulence, and magnetic fields.  A considerable amount of work has gone into any pair
of these, but relatively little study has combined all three with with high
resolution methods.  \citet{Mouschovias76}, \cite{Scott80}, and \cite{Galli93b}
have studied the combined effects of magnetic fields and gravity.
\citet{Goldreich95, Cho03} and \citet{Kritsuk09b} have studied the effects of
magnetic fields on turbulence, both compressible and incompressible.
Self-gravitating turbulence has been studied by several authors, notably
\citet{Klessen00} and \citet{Kritsuk11}, finding enhanced high density material
relative to the pure turbulence results.

Combining all three physical mechanisms, \citet{Price08}, \citet{Federrath11},
and \citet{Collins11} have done simulations of self-gravitating, magnetized
turbulence with high dynamic range methods. The primary difference in setup
between the first of those works \citep{Price08,Federrath11} and the work
presented here are initial and boundary conditions: in those works, the initial
conditions were isolated uniform density spheres with random velocity
perturbations; in the work presented here, we begin with fully developed
turbulence in a periodic domain.  Of course, both situations are idealized, with
the full nature of star formation potentially dependent on the molecular cloud
formation process as well.


In this work, we present three high resolution, high dynamic range simulations
of supersonic, \sa\ and \transa\ turbulence with self gravity.  We find that
gravity breaks the cloud into two distinct states, one low density turbulent
state and one high density collapsing state.  We discuss several of the dominant statistics used in
describing supersonic turbulence, the impact that the newly formed
self-gravitating state has on them, and the effect of magnetic fields.
In Section \ref{sec.Method} we discuss the numerical algorithms, initial
conditions, and simulation parameters.
In Section \ref{sec.Vrho} we discuss the density probability distribution function (PDF),
$V(\rho)$; followed by the density
power spectra $P(\rho,k)$ in Section \ref{sec.Prho}; the distribution of energy
in Section \ref{sec.Energy}; the magnetic probability distribution function in
Section \ref{sec.VB};  velocity and magnetic power spectra in Section
\ref{sec.Pv}, and finally the relative importance of compressible and solenoidal
modes in Section \ref{sec.HelmholtzBoth}.  We discuss possible implications on
star formation, and compare with recent observations, in Section
\ref{sec.discussion}. We summarize our findings in Section
\ref{sec.Conclusion}.

\section{Numerical Method, Simulations, Analysis}\label{sec.Method}

\hltwo{For the data presented in this paper, we solve the ideal MHD equations
with self-gravity using} the adaptive mesh refinement (AMR) code Enzo
\citep{Bryan95,O'Shea04} extended to MHD by \citet{Collins10}.  This code
uses the AMR algorithms developed by \citet{Berger89} and \citet{Balsara01}, the
hyperbolic solver of \citet{Li08a}, the isothermal HLLD Riemann solver
developed by \citet{Mignone07},  and the CT method of \citet{Gardiner05}.

\def\csunits{\ensuremath{c_{\rm{s},2}}}
\def\perrho{\ensuremath{n_{\rm{H},3}}}
\hltwo{We select the Mach 
number, $\mach$, virial parameter, $\avir$, and mean thermal-to-magnetic
pressure ratio, $\betao$ as
\begin{align}
    \mach &=\frac{v_{\rms}}{\cs}= 9\\
    \avir &= \frac{5 v_{\rms}^2}{3 G \rho_0 L_0^2} = 1\\
    \betao&=\frac{8 \pi \cs^2 \rho_0}{B_0^2} = 0.2, 2, 20,
\end{align}
where $v_{\rms}$ is the r.m.s. velocity fluctuation, $\cs$ is the sound speed,
$\rho_0$ is the mean density, $L_0$ is the size of the box, and $B_0$ is the
mean magnetic field.}

\hltwo{
These can be scaled to physical clouds as
\begin{align}
    \tff &= 1.1  \perrho^{-1/2} \rm{Myr}\\
    L_0 &= 4.6 \csunits \perrho^{-1/2} \rm{pc}\\
    v_{\rms} & = 1.8 \csunits \kms \\
    M &= 5900 \csunits \perrho^{-1/2} \msun\\
    B_0 &= (13, 4.4, 1.3) \csunits \perrho^{1/2} \mu \rm{G},
\end{align}
where $\csunits=0.2\kms$ and $\perrho=n_H/(1000 \percc)$ are the sound speed and
hydrogen number density, respectively, and we have used a mean molecular weight
of 2.3 amu per particle.}

\hltwo{This definition of $\avir$ is exact for a
uniform density sphere in isolation, and we have used it here for consistency with other
works in the literature.  In reality, 
the actual importance of gravity relative to
kinetic energy difficult to ascertain for a turbulent box with periodic
boundaries due to the infinite nature of the box and the intermittent nature of
dense structures.  The value of $\avir$ used here is perhaps somewhat lower than the
average value from observed molecular clouds \citep{Heyer09, Dobbs11}, but not
outside the observed parameter range.  Furthermore, the virial parameter is
potentially scale-dependant, with larger clouds being on the average more
gravitationally bound than smaller ones \citep{Heyer01, Goodman09}, so it is possible that
these results apply better to a subset of large, gravitationally bound molecular
clouds.  }

The initial conditions for this simulation were generated by a suite of unigrid
simulations using the PPML code \citep{Ustyugov09} without self-gravity.  Cubes
with $1024^3$ zones, with initially uniform density and magnetic fields, were
driven using a solenoidal driving pattern.   Power in the driving was between
wavenumbers $\kkmin=1,2$, and driven as in \citet{MacLow99} to maintain our
target Mach number.
Driving continued for
several dynamical times, 
\begin{align}
    \hltwo{\tdyn=\frac{L_0}{2 v_{\rms}}=1.2 \perrho^{-1/2} \rm{Myr}.}
\end{align}
until a statistically relaxed state was reached.  Results of the
turbulent boxes were first presented in \citet{Kritsuk09b}, see
Kritsuk et al. (2012, in preparation)
for more details.

\begin{figure*} \begin{center}
\ifpdf
\includegraphics[width=\hw\textwidth]{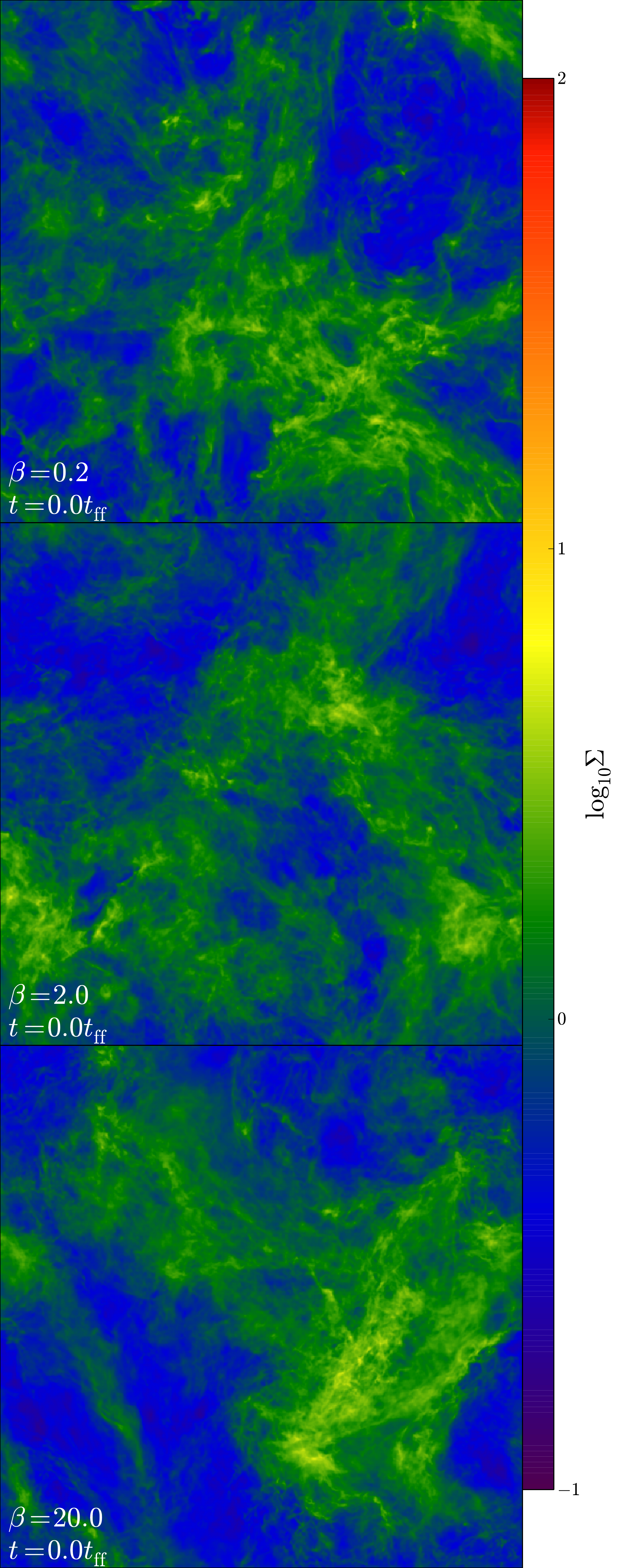}
\includegraphics[width=\hw\textwidth]{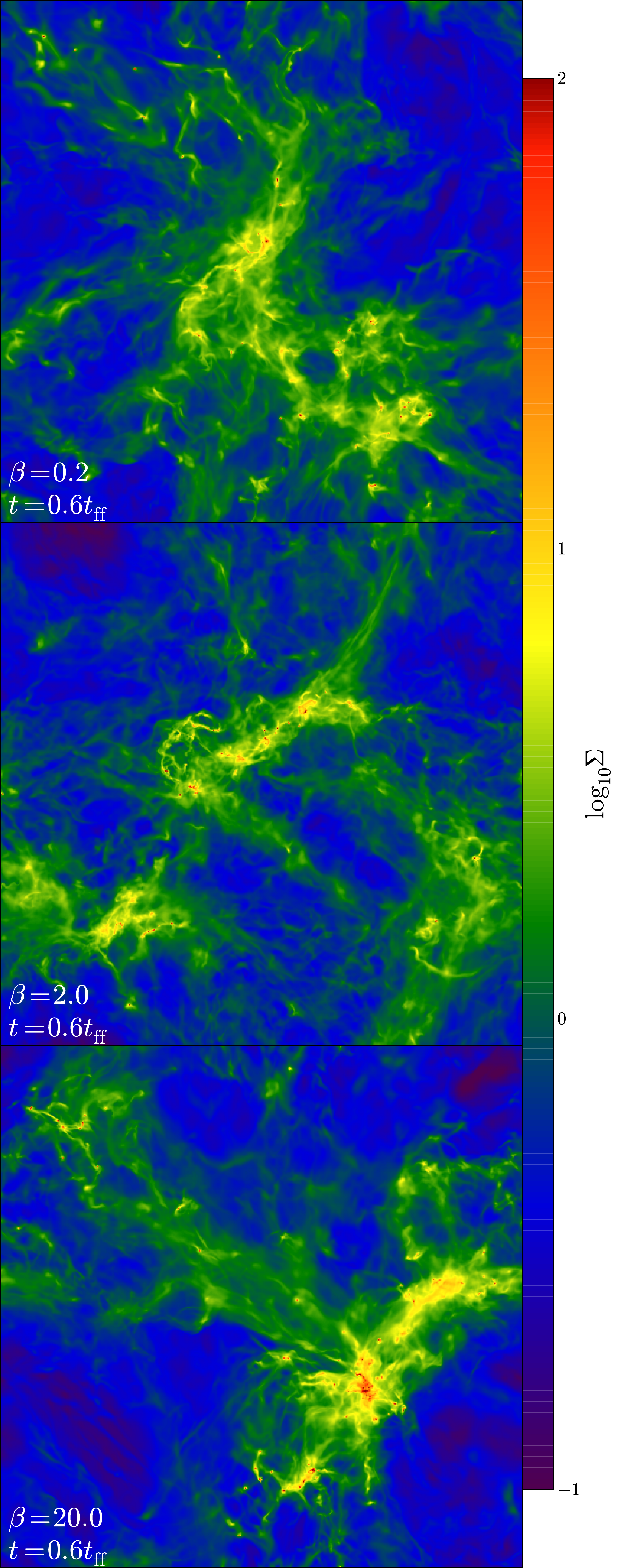}
\else
\includegraphics[width=\hw\textwidth]{figs_images/Proj_0000b025vb25vb205v.eps}
\includegraphics[width=\hw\textwidth]{figs_images/Proj_0060b025vb25vb205v.eps}
\fi
\caption[ ]{ Column density ($\Sigma$) for the two primary snapshots ($t=0.0\tff$, left
column, and \tlate, right column), for all three simulations
(from top to bottom, \betal, 2, 20, respectively).  Both figures show the
filamentary structure expected from both turbulence models and observations.
The right column shows the addition of high density collapsing gas, and somewhat
enhanced filamentary nature.}
\label{fig.projections} \end{center} \end{figure*}

The simulations were then restarted using Enzo with self-gravity and AMR.
A root grid of $512^3$ and 4 levels of refinement by a factor of 2 were
used, with refinement such that the local Jeans length $L_{\rm{J}} = \sqrt{
\cs^2 \pi/G \rho}$ is resolved by at least 16 zones.  This gives an effective
linear resolution of $8,192$.  The simulations were run for $0.6\tff$.
Figure
\ref{fig.projections} shows projections through the volume at $t=0.0$ (left
column) and $t=0.6\tff$ (right column) for each \betao\ (top to bottom, \betal,
2, and 20).  In the left column, one can see the filamentary structures
associated with supersonic turbulence.  The right column shows the distribution
of high density collapsing cores superimposed on the turbulent state.

It should be noted that the two solvers used for this simulation have different
dissipation properties.  The solver used for the initial conditions (PPML)
employs a third order spatial reconstruction, while the solver used in the
self-gravitating portion was only second order spatially.  \hltwo{This change in
solver was due to the fact that AMR as employed in Enzo has not yet been extended to include the
increased algorithmic complexities of the higher quality PPML algorithm.}
Details about the differences in numerical dissipation can be found in
\citet{Kritsuk11b}.  In that work, the solver used in these AMR runs is referred
to as LL-MHD.  The effects of this solver change can be seen most prominently as
a minor loss of dynamic range in the velocity power spectra, but the statistical
properties are otherwise the same between the two solvers.

In order to demonstrate the effects of gravity in the rest of the paper, we
typically present the simulations at two fiducial times, $t=0.1\tff$ and
$0.6\tff$. The first snapshot was taken at $t=0.1\tff$, which is sufficient to
remove the effects of the solver transition, but it is early enough to not show
any effects of gravity.  Unless otherwise noted (e.g., velocity power spectra,
Figure \ref{fig.PvFull}) the statistics at $t=0.1\tff$ are identical to those at
$t=0.0\tff$, which corresponds to the moment at which gravity was turned on.
Due to the short timescale relevant for the high density gas, we average several
snapshots around each of the two fiducial snapshots, in a range of $\pm
0.05\tff$, to reduce statistical noise.  This short-term time averaging is done
in all plots unless otherwise noted.

\def\Nc{\ensuremath{\mathcal{N}}}
\def\tsim{\ensuremath{t_{\rm{sim}}}}
\def\tl{\ensuremath{t_{\ell}}}
\def\zp{\ensuremath{\zeta_{\rm{p}}}}
\def\pp{\ensuremath{\rm{p}}}

The final time, $\tsim=0.6\tff$, corresponds to $0.5\tdyn$.  
We can estimate the scale at which the turbulence can be considered relaxed by
using the structure function scaling, $\delta v_\ell^p \propto \ell^{\zp}$, and
computing the scale at which the number of turnovers at the end of our
simulation is greater than some number, $\Nc = \tsim/ \tl $,
were $\tl=\ell/\delta v_\ell$ is the turnover time
at length $\ell$.  One finds
\begin{align}
\ell_\Nc = \left( \frac{V_0 \tsim}{\Nc}
\right)^{\pp/(\pp-\zp)}L_0^{\zp/\left(\pp-\zp\right)},
\end{align}
where $V_0=10$ is the velocity at the outer scale, $L_0=1$ is the size of the
box, and $\Nc$ is the number of crossings at a given scale.  For the third order
structure function, $\pp=3$, in a supersonic flow, \citet{Kritsuk07} found that
$\zeta_{3}=1.3$.  For a single turnover time, $\Nc=1$, we
find $\ell_3=0.09$, which corresponds to $\kkmin=11$.  The short time averaging
window, $\pm0.05 \tff$, corresponds to $\kkmin=875$, which is resolved by all
refined regions.

Power spectra in this simulation were computed only using the root grid data, at
$512^3$.  Due to the incomplete filling of $k$-space, power spectra that also
include the refined regions would require data interpolation.  \hlthree{Due to
the fact that the volume filling fraction of refined regions is quite low (see
Section \ref{sec.Vrho}), and the fact that power spectra are volume weighted
quantities, spectra using anything but the root grid data would be dominated by
interpolated data that do not contain much useful information. }

All analysis has been performed with
the AMR analysis package {\tt yt} \citep{Turk11}.

\section{Density PDF}\label{sec.Vrho}
Figure \ref{fig.VrhoBoth} shows the density
PDF, $V(\rho)$, for each simulation (\betal\ in red, \betam\ in green, and \betah\ in
blue) at \tearly\ (solid lines) and \tlate\ (dotted lines).  This color scheme will
be used throughout the paper.  The figure additionally shows two grey lines that
divide the gas into three sections; low density turbulent gas (left section),
high density self-gravitating gas (center section), and very high density gas that is numerically
unresolved (right section).  The
first two states are of the greatest interest to us here.  The last state is
interesting qualitatively, but we cannot make any quantitative measurements of
the gas here.  \hlthree{The transition density between turbulent and collapsing
states, $\rholow$, is taken
where the power-law begins to transition from 
lognormal, at $\rholow=10$.}  As
there is likely shock-compressed gas
above $\rholow$ that is not self-gravitating, $\rholow$ is not meant to be used
as a phase boundary between the two states.  \hlthree{A more complete set of criteria for
the transition between turbulent and collapsing gas is currently under
investigation.}
The second division is taken at
the highest density that is still considered resolved by our refinement
criterion on the finest level, as discussed in Section \ref{sec.Method}. This gives
$\rhohigh=6347$.  At this density there is also a change in the power-law
slope.  At very high density, corresponding to very small scale, inaccuracies in
the angular momentum transport become dominant and the gas cannot collapse
fully, leading to excess mass and decreased fragmentation.  This can be somewhat
addressed by incorporating sink particles, but as of this writing no
satisfactory prescription of sink particles with magnetic fields has been
developed.

In the following, we will identify the turbulent state as those features that
belong to either low density gas (below $\rholow$) or gas that whose statistical
properties are relatively unchanged over the course of our simulation.
Collapsing gas is identified by high density (between $\rholow$ and $\rhohigh$) and short time variation.

\subsection{Density PDF in the Turbulent State}

One of the most robust properties of isothermal supersonic turbulence is the
lognormal distribution of densities
\citep{Blaisdell93, Vazquez-Semadeni94,
Padoan97a, Padoan97b,
Scalo98,Passot98,Nordlund99,Kritsuk07, Federrath08,Price12}.
Several properties of star formation have been predicted using the lognormal distribution
function,  including
the initial mass function (IMF) of stars
\citep{Padoan02,Padoan07}, brown dwarf frequency \citep{Padoan04} and the star
formation rate
\citep{Krumholz05, Padoan11}. 

\hlthree{For compressible turbulence without gravity or magnetic fields},  the density PDF,
$V(\rho)$, can be shown to be  a lognormal of the form
\begin{align}
V(\rho) d\ln \rho = \frac{1}{\sqrt{2 \pi \sigma^2} }~ \rm{exp}\left( \frac{( \ln \rho -
    \mu)^2}{2 \sigma^2} \right ) d\ln \rho,
\label{eqn.lognormal}
\end{align}
where $\mu =
-\sigma^2/2$ is the mean of $\ln \rho$ \citep{Blaisdell93,Vazquez-Semadeni94}.
The variance and Mach number, \mach, are related by
\begin{align}
\sigma = \sqrt{\ln(1+b^2 \mathcal{M}^2)}.
\label{eqn.lognormalWidth}
\end{align}
The parameter $b$ has been determined numerically to lie between 0.3 and 0.4
\citep{Padoan97b,Federrath08, Kritsuk07, Beetz08, Kritsuk09a, Federrath10}. The
value $b$ has been shown to approach unity in simulations with compressive
forcing \citep{Federrath08, Federrath10}.

\hlthree{The presence of the Lorentz terms in the momentum equation breaks
the invariance of the equations relative to the mean density.  Thus the
sequence of shocks that determine the density of a parcel of gas are no longer
independent multiplicative events, as they are in hydrodynamic turbulence.  For
this reason there is no a priori expectation of a
lognormal density distribution in a magnetized system.}
However, \citet{Ostriker01} and \citet{Lemaster08} have shown that
$V(\rho)$ is approximately lognormal, with properties weakly depending on mean
field strength. 
For driven MHD turbulence, \citet{Lemaster08} find that, for densities within
$10\%$ of the peak density,
\begin{align}
\sigma_{\rm{LS08}} = \sqrt{|-0.72 \ln\left(1+0.5 \mathcal{M}^2\right) + 0.20|},
\end{align}
and weakly decreases with field strength.  Kritsuk et al. (2012, in
preparation)
performed
high resolution simulations 
of statistically stationary magnetized turbulence and showed that the presence
of a magnetic field alters the low density gas, making a lognormal description less
appropriate. They did
find that the high density wing of the PDF is still well
approximated by a lognormal for the two weak field runs, $\betam$ and $\betah$.

Figure \ref{fig.VrhoBoth} shows the resemblance to a lognormal present in
our simulations.  In Table \ref{table.lognormal_fits}, we show $\mu$ and
$\sigma$ found from fits to the average PDF for three snapshots from each of our
simulations.  Fits were performed for $\rho\in[5\sci{-3},10]$, a range chosen
to exclude material \hlthree{above $\rholow$.}
We find a weak sensitivity of
$\mu$ and $\sigma$ with both time and $\betao$, with $\mu$ decreasing with time
and increasing with $\betao$.  Our short averaging time, relative to a dynamical time,
means that these can be viewed only as snapshots in time, \hltwo{not as robust
statistical averages.}

\subsection{$V(\rho)$ Collapsing State}\label{sec.VrhoCollapse}
The PDF of the collapsing material forms 
a power-law, $V(\rho)\propto \rho^\vrex$, \hlthree{for densities above
$\rholow$.}  This was
first presented in hydrodynamic simulations by \citet{Klessen00}, though the resolution of those simulations
was too low to measure a significant power-law.  The first measurement of the slope
was done by \citet{Slyz05}, who found $\vrex=-1.5$.  This slope
was also seen by \citet{Federrath08b} and \citet{Vazquez08}.  
 \citet{Kritsuk11} used a very high resolution AMR simulation to measure a slope
of $-1.67$ at intermediate to high densities, and $-1.5$ at high density.  They
also provided and explanation of this power-law 
by comparing $V(\rho)$ to that of self-similar
collapse, wherein $\rho\propto r^\rrex$.  
The three solutions they discussed are the
pressure free collapse \citep[PF]{Penston69},
Larson-Penston supersonic infall \citep[LP]{Larson69,Penston69}, and expansion wave
from inside out collapse \citep[EW]{Shu77}.  Values of $\rrex$ for various models, and the implied
value of $\vrex$, are shown in Table \ref{table.self_similar_exponents}.  This
table summarizes all semi-analytic and numerical results in this paper.
\citet{Collins11} measured this exponent for \sa\ MHD turbulence, and found a
value of \vrex=$-1.64$ for $\rho\in[10,1000]$.  For the simulations in this work,
we find $\vrex = -1.80,\ -1.78,$ and $-1.65$ for $\betal,\ 2$, and $20$.  The
\betah\ case presented here is similar in both physical parameters and measured
slope to that in \citet{Collins11} and in the gas dynamic simulations of
\citet{Kritsuk11}, while the stronger field simulations show
steeper values.  The values found here are most consistent with the
pressure-free value of \hlthree{$\vrex=-1.75$}.  

\begin{figure} \begin{center}
\ifpdf
\includegraphics[width=\hw\textwidth]{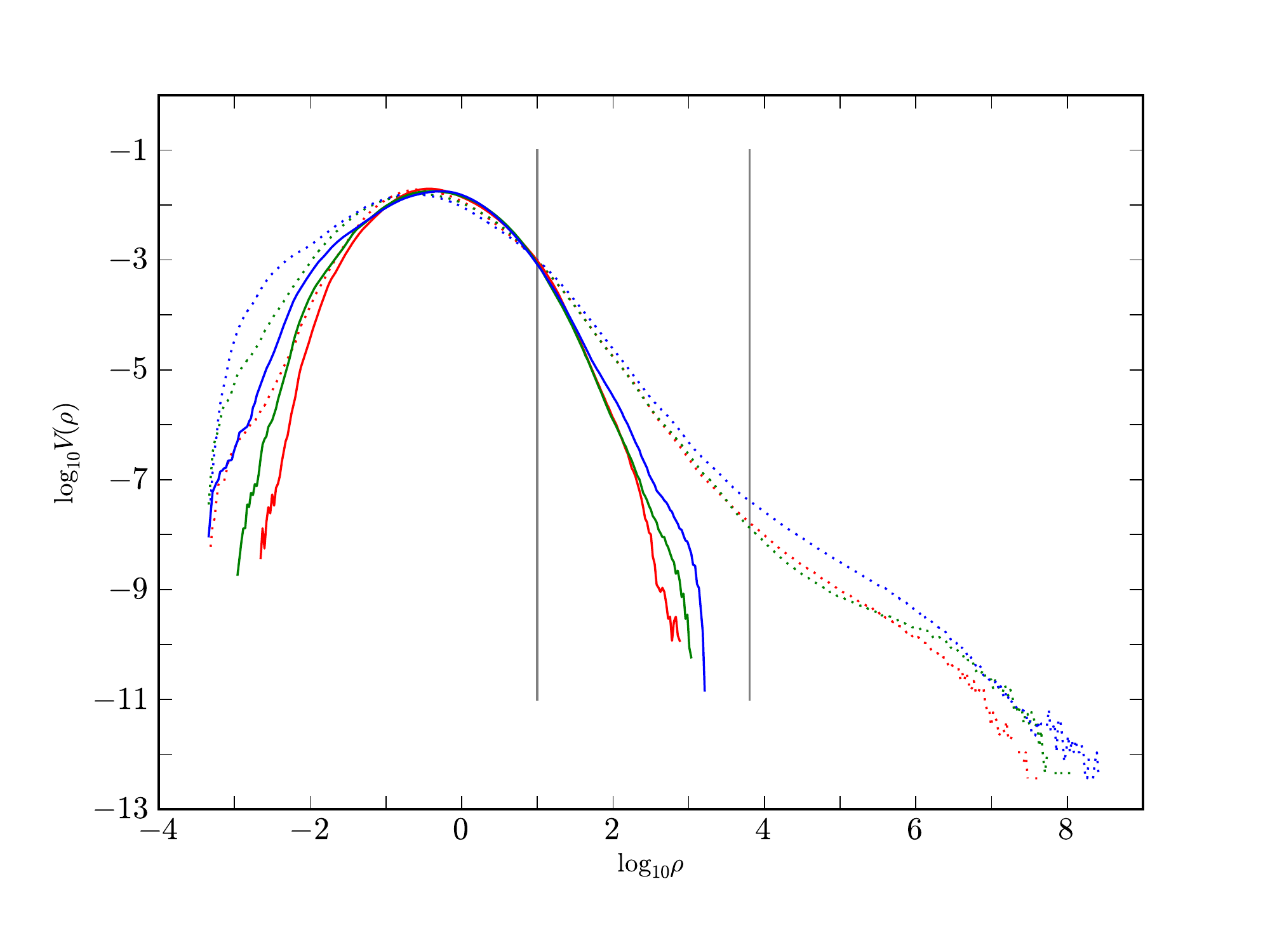}
\else
\includegraphics[width=\hw\textwidth]{figs_PDF//DensityVolume.eps}
\fi
\caption[ ]{Density PDF, $V(\rho)$, for each of our three-simulations, 
$\beta_0=0.2,\ 2$, and $20$ (red, green, and blue respectively),  at two
snapshots,
$t=0.1,0.6 \tff$ (solid and dotted lines, respectively).  The vertical grey
lines
separate the low density turbulent state (left section), high density
collapsing state (center section) and very high density unresolved gas (right
section).}
\label{fig.VrhoBoth} \end{center} \end{figure}

\begin{figure} \begin{center}
\ifpdf
\includegraphics[width=\hw\textwidth]{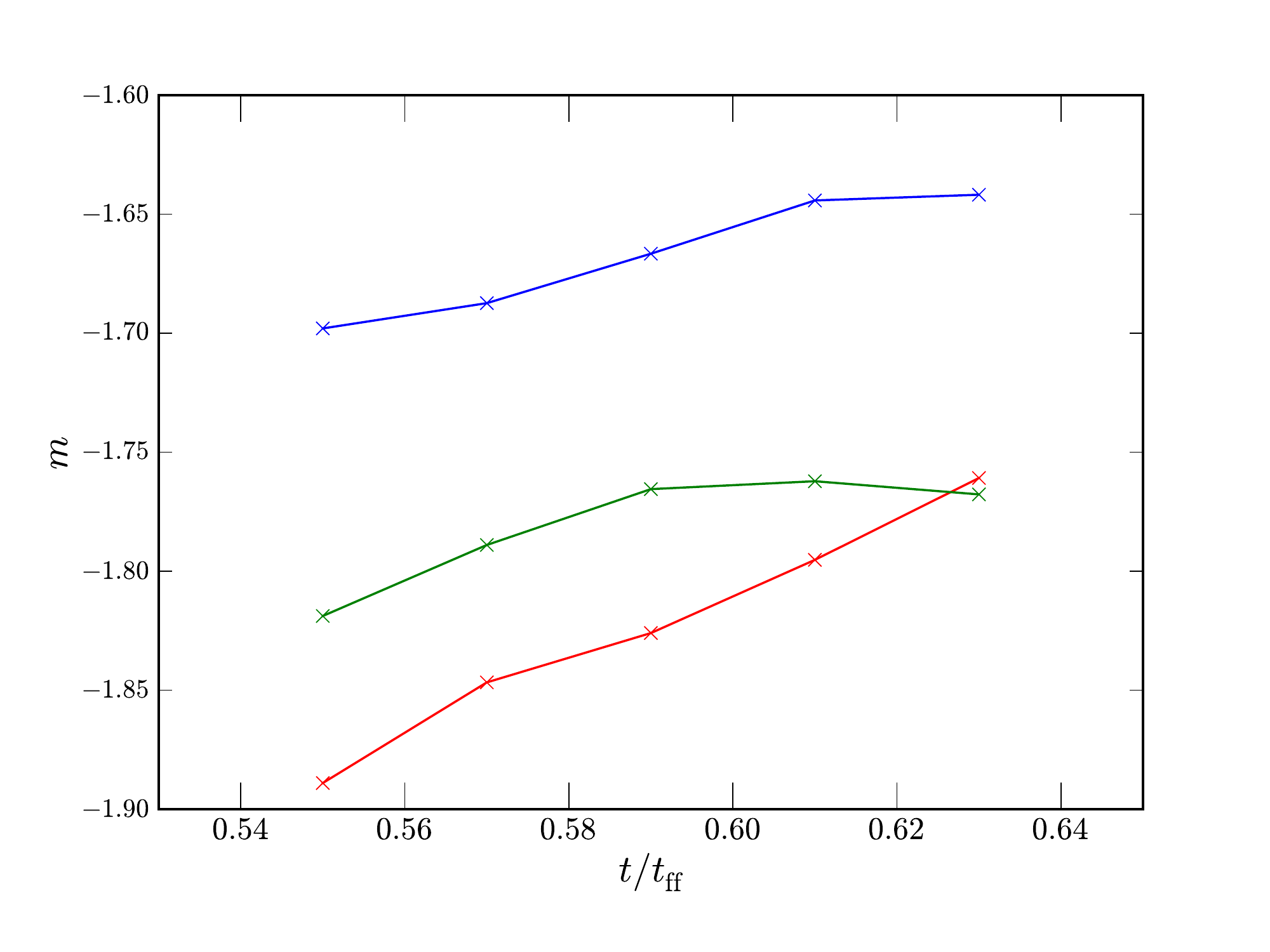}
\else
\includegraphics[width=\hw\textwidth]{figs_PDF//vrex_convergence.eps}
\fi
\caption[ ]{The evolution of the density PDF power-law exponent, $\vrex$, for
the 
$\beta_0=0.2,\ 2,$ and $ 20$ simulations (red, green, and blue respectively).
The value of this exponent
seems to be increasing for the \betal\ simulation, but possibly reaching a
constant value for
the other two.}
\label{fig.vrex_convergence} \end{center} \end{figure}

It is clear from Figure \ref{fig.VrhoBoth} that the slope is a function of
time in these clouds.  
This was also
discussed by \citet{Kritsuk11}. 
Figure \ref{fig.vrex_convergence} shows the time evolution of the power-law
exponent $\vrex$ for the last few snapshots, those that contributed to the
average shown by the dashed curve in Figure \ref{fig.VrhoBoth}.    The two low field cases, $\betao=2$ and $20$,
seem to have converged, with power-law index $\vrex$ in the $\betam$ case
slightly lower than the
$\betah$ simulation.  The $\betal$ simulation has clearly not converged, and is
still increasing at the end of the simulation.
Given the agreement of $\vrex\approx-1.64$ between our $\betah$ simulation and
previous simulations \citep{Collins11,Kritsuk11} and observations, we feel confident that this is a
robust result in the hydrodynamic limit.

As we will discuss in Sections 
\ref{sec.HelmholtzBoth} and \ref{sec.WhereIPutSFR}, three
dimensional compressions are suppressed with increasing magnetic field strength,
as the flow is forced along magnetic field lines.  This reduction in
compressibility is likely the reason for the reduced value of $\vrex$ in
$\betam$, and delayed convergence in the $\betal$ run.

\begin{figure} \begin{center}
\ifpdf
\includegraphics[width=\hw\textwidth]{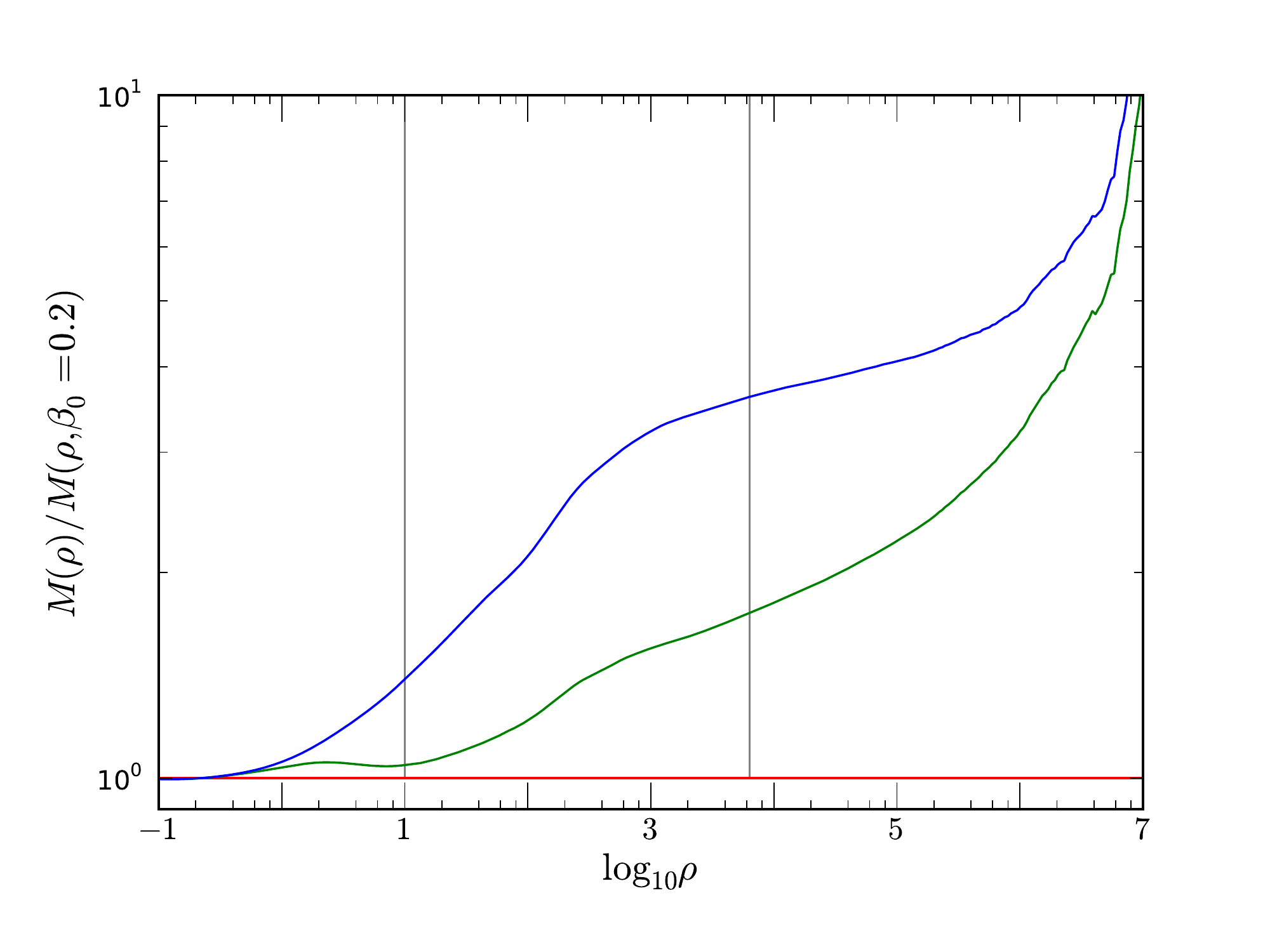}
\else
\includegraphics[width=\hw\textwidth]{figs_PDF//MgRelativeb025vb25vb205vGreatMassFraction_63.eps}
\fi
\caption[ ]{Cumulative mass relative to $\betal$\ case for the last snapshot,
$t=0.6\tff$. Vertical lines separate turbulent (left) collapsing (center) and
unresolved (right) gas.}
\label{fig.GreatMassRelative} \end{center} \end{figure}

\begin{figure} \begin{center}
\ifpdf
\includegraphics[width=\hw\textwidth]{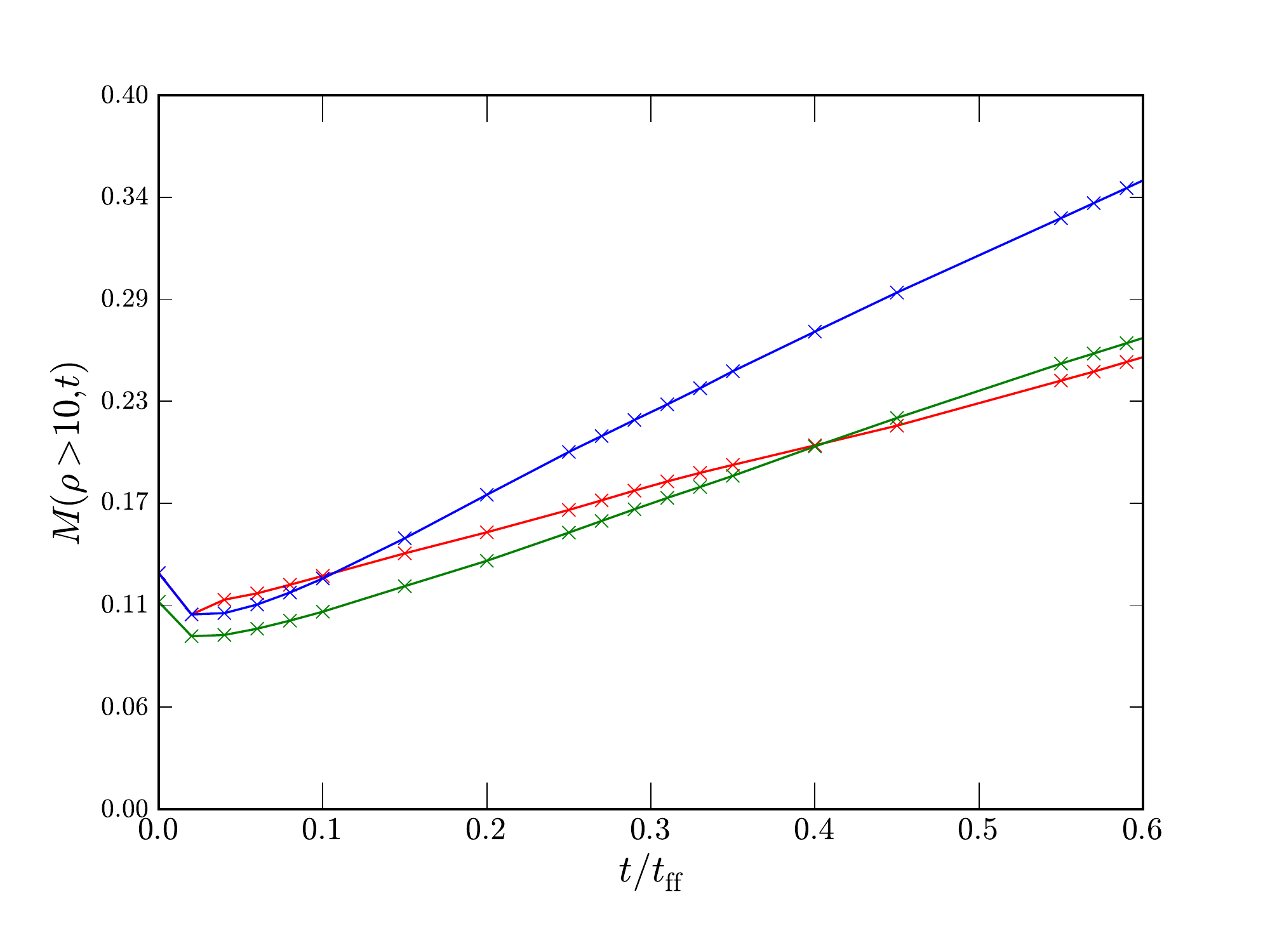}
\else
\includegraphics[width=\hw\textwidth]{figs_rate//SFR_512.eps}
\fi
\caption[ ]{Cumulative mass for $\rho>10$  vs. time for $\betal, 2$, and $20$ (red, green, and
blue, resepctively). This shows the rate at which gas enters the collapsing
state, and is a decreasing function of $\betao$.}
\label{fig.sfr} \end{center} \end{figure}

The cumulative mass fraction above some critical density $\rhoc$, defined as 
$$
M(\rhoc)=\int_{\rhoc}^\infty \rho V(\rho) d \rho,
$$
has been used in a number
of theories of star formation, as we will discuss in Section \ref{sec.theory}.
In order to quantify the effects of mean field, Figure
\ref{fig.GreatMassRelative} shows $M(\rhot)$
for each run relative to the \betal\ simulation.  This shows an increase in
the collapsed mass as a function of field strength.  The behavior of
this mass fraction with time, $M(\rho,t)$, can be used to
explain the slower convergence of $\vrex$ with the more magnetized simulations.
Figure \ref{fig.sfr} shows $M(\rhoc=10,t)$, which shows that the rate at which
material enters the high density state is a decreasing function of $\betao$.
This can be used as a proxy for star formation, as we will discuss in Section
\ref{sec.theory}, and while the exact rate is a function of the critical density
$\rho_{\rm{c}}$, the increase of rate with \betao\ is not.

\begin{table}[h]
\begin{center}
\caption{Lognormal Fit Parameters}
\label{table.lognormal_fits}
\begin{tabular}{|r|r|r|r||r|r|r|}
\hline
&  \multicolumn{3}{|c||}{$\sigma$} & \multicolumn{3}{|c|}{ $\mu$} \\
\hline
\backslashbox{time}{$\betao$} &0.2 & 2.0 & 20 & 0.2 & 2.0 & 20\\
\hline
0.1 & 1.2 & 1.3 &1.3&  $-0.86$ &$-0.84 $&$-0.80$ \\
0.3 & 1.2 & 1.4 &1.4&  $-0.97$ &$-1.00 $&$-1.1 $\\
0.6 & 1.2 & 1.5 &1.5&  $-1.1 $ &$-1.2  $&$-1.4 $\\
\hline
\end{tabular}
\end{center}
\end{table}

%
%
%
%

\ctable[
sideways,
star,  caption = {Power-law relationships, predictions, and
measurements.},label={table.self_similar_exponents}]{l c c c c c c c c c }{\tnote[a]{Measured values of \rrex\ computed from measured \vrex.}
\tnote[b]{Semi-analytic prediction of \vbex\ uses q=1/2.}
\tnote[c]{Values for early and late times, respectively.  Late times contain information from both turbulent and collapsing states.}
\tnote[d]{Values for turbulent state only.}
\tnote[e]{Values for low and high densities, respectively}
}{\hline
\hline
& $\rho\propto r^{\rrex}$& $v\propto r^\urex$& $V(\rho)\propto \rho^\vrex$& $V(B)\propto B^\vbex$& $P(\rho,k) \propto k^\prex$ & $P(\Sigma,k)\propto k^\psigmaex$& $P(v,k)\propto k^\pvex$& $B \propto \rho^\brex$& $\beta \propto \rho^\betarex $\\ 
latex& rrex = \rrex& urex=\urex& vrex=\vrex& vbex=\vbex& prex=\prex& psigmaex = \psigmaex& pvex = \pvex& brex = \brex& betarex = \betarex\\ 
Index& \rrex& \urex& \vrex = 3/\rrex& \vbex = \vrex/\brex& $\prex = -2(\rrex+1)$& $-2(\rrex+2)$& $\pvex=-2(\urex+1)$& \brex& $\betarex = 1-2\brex$\\ 
\hline
LP& $-2$& $0$& $-1.5$& $-3$\tmark[b]& $2$& $0$& $\delta(k)$& ...& ...\\ 
PF& $-12/7$& $1/7$& $-1.75$& $-3.5$\tmark[b]& $10/7$& $-4/7$& $9/7$& ...& ...\\ 
EW& $-3/2$& $-1/2$& $-2$& $-4$\tmark[b]& $1$& $-1$& $0$& ...& ...\\ 
Isotropic& ...& ...& ...& ...& ...& ...& ...& $2/3$& $-1/3$\\ 
$\vvec\parallel\bvec$& ...& ...& ...& ...& ...& ...& ...& $0$& $1$\\ 
$\vvec\perp\bvec$& ...& ...& ...& ...& ...& ...& ...& $1$& $-1$\\ 
\hline
$\beta=0.2$& $-1.67$\tmark[a]& ...& $-1.80$& $-5.4$& $-0.42,\ 0.86$\tmark[c]& $-1.39,-0.11$\tmark[c]& $-1.46$\tmark[d]& $0.02,\ 0.43$\tmark[e]& $0.96,\ 0.15$\tmark[e]\\ 
$\beta=2$& $-1.69$\tmark[a]& ...& $-1.78$& $-4.0$& $-0.58,\ 1.12$\tmark[c]& $-1.61,0.15$\tmark[c]& $-1.58$\tmark[d]& $0.12,\ 0.40$\tmark[e]& $0.77,\ 0.19$\tmark[e]\\ 
$\beta=20$& $-1.82$\tmark[a]& ...& $-1.65$& $-3.2$& $-0.62,\ 1.2$ \tmark[c]& $-1.53,0.25$\tmark[c]& $-1.80$\tmark[d]& $0.23,\ 0.39$\tmark[e]& $0.54,\ 0.23$\tmark[e]\\ 
\hline
}

\section{Density Power Spectra}\label{sec.Prho}

\begin{figure} \begin{center}
\ifpdf
\includegraphics[width=\hw\textwidth]{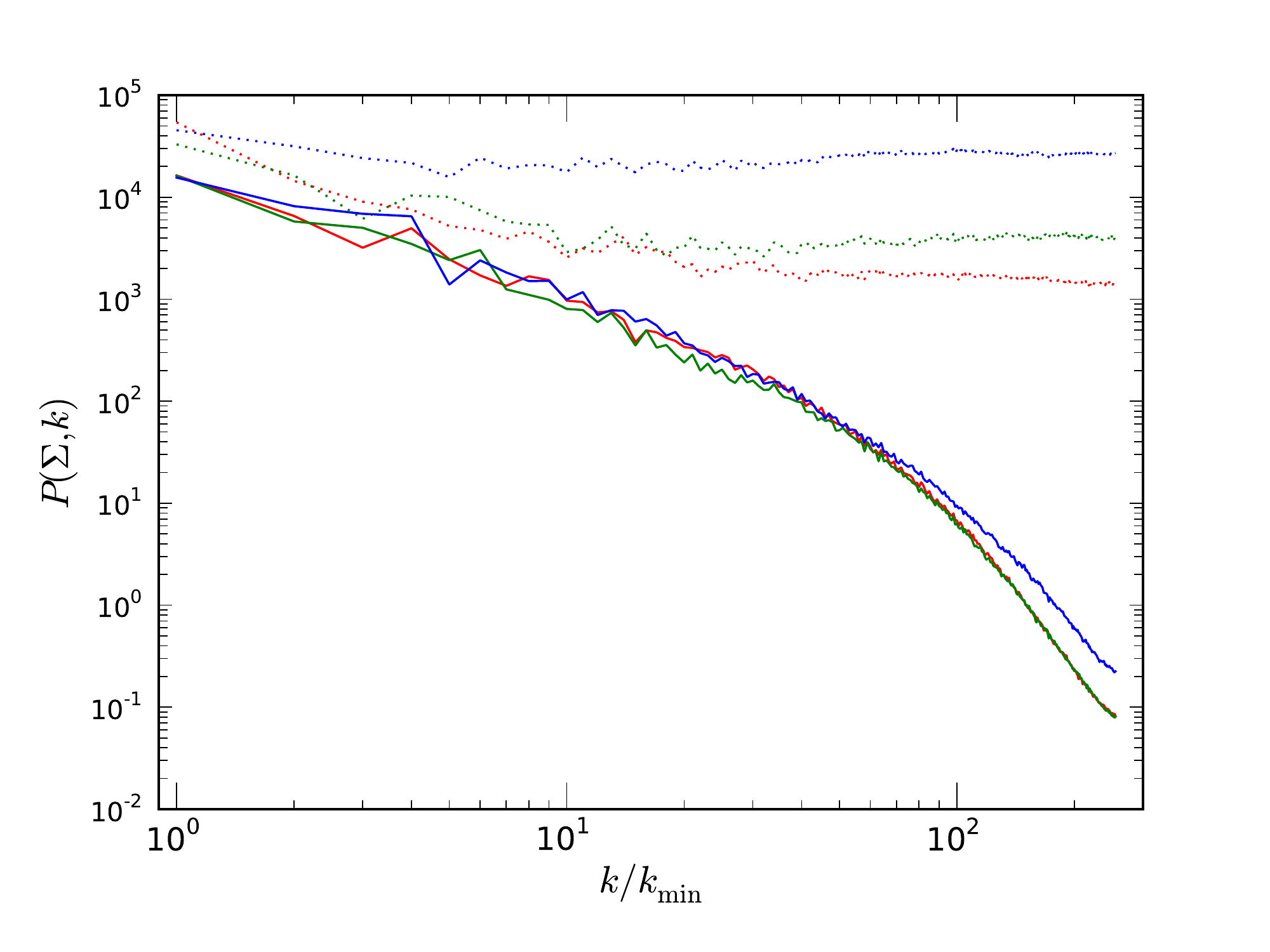}
\else
\includegraphics[width=\hw\textwidth]{figs_spectra//spectra_rhob025vb25vb205vprojection_Density_2_4_6_8_10_57_59_61_63_65.eps}
\fi
\caption[ ]{Column density power spectra, $P(\Sigma,k)$.  Plot style is the same
as in
Figure \ref{fig.VrhoBoth} }
\label{fig.PSigmaBoth} \end{center} \end{figure}

One of the more striking results from this study is the behavior of the density
power spectrum, $P(\rho,k)$, under the effects of self-gravity.  Here we have
defined $P(\rho,k)$ as
\begin{align}
P(\rho,k) = \sum_{|\p{k}|=k} \tilde{\rho}_{\p{k}}^* \tilde{\rho}_{\p{k}},
\end{align}
where $\tilde{\rho}$ is the Fourier transform of $\rho$, and the star denotes
its conjugate.  Figure \ref{fig.PrhoBoth} shows the density power spectra,
$P(\rho,k)$, for all three simulations ($\betao=0.2,\ 2,$ and $20$ in  red,
green, and blue, 
respectively) at two snapshots, \tearly\ and \tlate\ (solid, dotted).  
In Figure \ref{fig.PlnRhoAll} we show the power spectra for $\ln \rho$, for three
snapshots of the $\betal$ simulation. 
Figure \ref{fig.PSigmaBoth} shows the column density power
spectra, $P(\Sigma,k)$.  In Figures \ref{fig.PrhoBoth} and
\ref{fig.PSigmaBoth}, the birth of the high density collapsing state can be seen
in a dramatic change in the behavior of of $P(\rho,k)$, transitioning to a
positive slope.  This behavior is conspicuously
absent in the power spectrum of the logarithm of density, as we will discuss in
the following sections.

\subsection{Density Power Spectra in the Turbulent State}\label{sec.PrhoTurb}
The turbulent initial conditions can be seen in the early snapshot (solid line)
in Figure \ref{fig.PrhoBoth}.  For weakly subsonic (but still compressible)
isothermal
turbulence, one expects the density to follow the pressure fluctuations, and the
density power spectrum, 
$P(\rho,k) \propto
k^\prex$, should scale as $\prex=-7/3$ \citep{Bayly92,Kritsuk07}.   
It has been seen that this power spectrum flattens with increasing Mach number.
For trans-sonic
turbulence, \citet{Kim05} measured $\prex=-1.7$ at a Mach number $\mach=1.2$.
For supersonic turbulence, with $\mach=6$, \citet{Kritsuk07} measure
$\prex=-0.9$ for a simulation with $512^3$ zones, and a somewhat steeper
spectrum of $\prex=-1.07$ for $1024^3$.  For the early snapshot, we measure
significantly more shallow spectra, $\prex=-0.42, -0.58,$ and $-0.62$ for our
$\betal$, 2, and 20 simulations, respectively.  This fit was done for
$\kkmin\in[2,30]$.  The flatter nature of these
spectra, relative to the \citet{Kritsuk07} work, is likely due to the higher
Mach number employed in our simulations.  For Burger's equation with vanishing
pressure term, which could be analogous to increasingly supersonic turbulence,
\citet{Saichev96} predict that $\prex=0$, a trend that is consistent with
$\prex$ here being shallower than that found in \citet{Kritsuk07}.
It should be noted that this comparison to Burger's turbulence is not meant to imply that supersonic
turbulence is similar to turbulence in  Burger's equation; the lack of vorticity
and strong intermittency in Burger's equation makes the two systems only
superficially similar.  A second potential cause of the change in slope is due
to the fact that the solver employed here is more diffusive that PPM used in
\citet{Kritsuk07, Kritsuk11b}, which may contribute to the change in slope.  The
decrease in slope with increasing magnetic field strength is consistent with the
decrease in slope in the velocity power spectrum, 
as will be discussed
in Section \ref{sec.Pv}.  

\begin{figure} \begin{center}
\ifpdf
\includegraphics[width=\hw\textwidth]{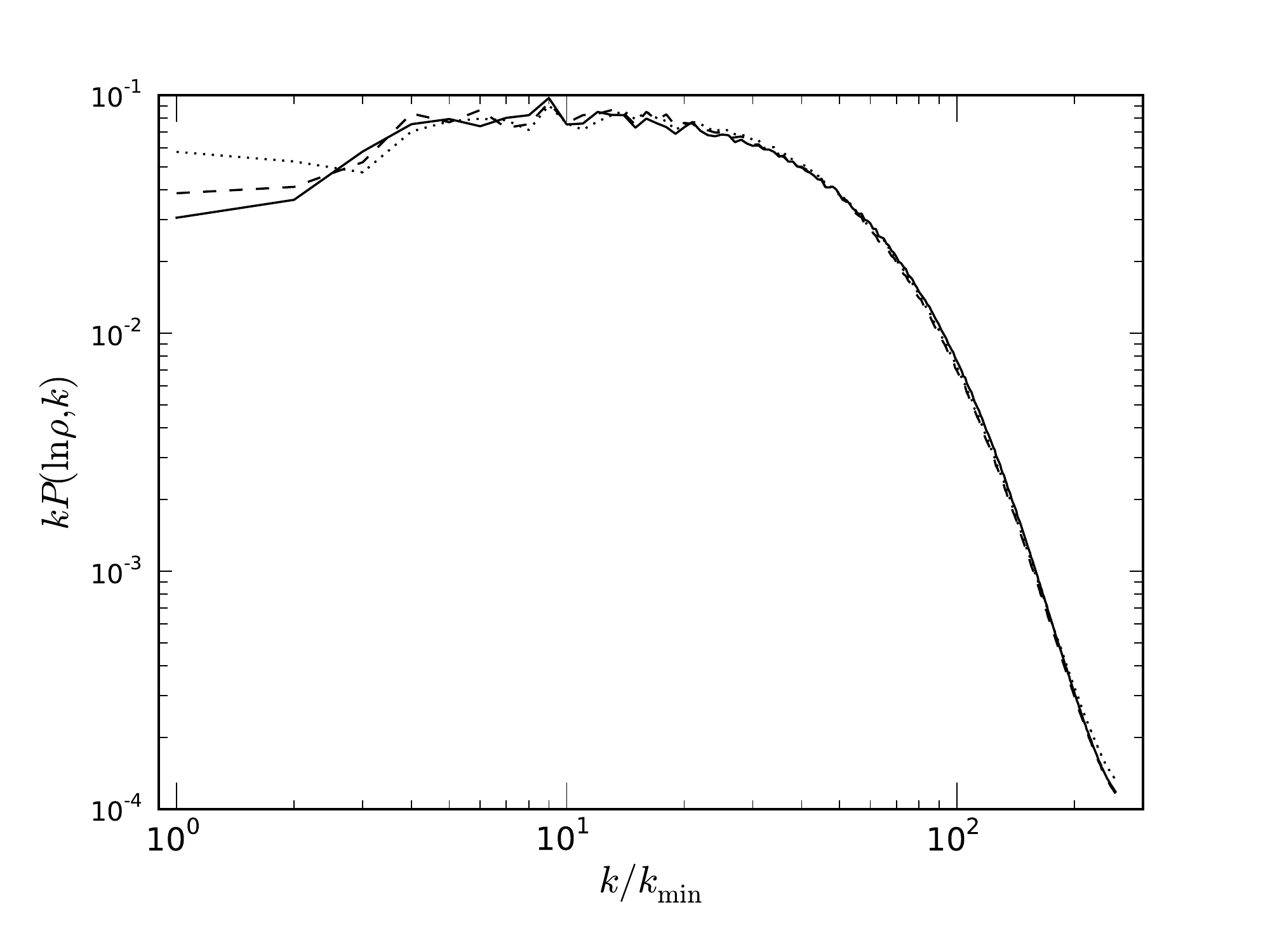}
\else
\includegraphics[width=\hw\textwidth]{figs_spectra//spectra_lnrhob025vLogDensity_10_35_63.eps}
\fi
\caption[ ]{Power spectra for the logarithm of density, $P(\ln\rho,k)$, for
\tearly, \tmid, and \tlate (solid, dashed, and dotted lines) for the $\betal$
simulation.  Taking the spectrum of the logarithm of density allows us to
recover the turbulent state even at late times, as the collapsing state is
filtered out by the logarithm. Slopes get somewhat flatter with increasing mean
field; the $\betal$ simulation has a slope of $-1.0$, and the $\betam$ and $20$
simulations having
slopes of $-1.4$ and $-1.53$, respectively (neither shown here)}
\label{fig.PlnRhoAll} \end{center} \end{figure}

As discussed in \citet{Beresnyak05}, $P(\rho,k)$ is strongly influenced by the
rare, high density peaks.  In that work, the authors noticed that the
power-spectrum of the logarithm of density was significantly steeper than
$P(\rho,k)$, as the
logarithm effectively filters out high density, rare peaks.  This is seen in our
simulations as well.  The behavior of $P(\ln \rho,k)$ shows very little
evolution with time, as the collapsing state is effectively filtered out, which
allows us to recover the turbulent state through the entire simulation.  This
can be seen in Figure \ref{fig.PlnRhoAll} for the $\betal$ simulation.  Slopes
get somewhat steeper with increasing mean field, with $\betal$ having a slope of
$-1.0$, and $\betam$ and $20$ having slopes of $-1.4$ and $-1.53$, respectively
(neither shown here).  


In order to make contact with observable quantities, we show the power spectra
for column density $\Sigma$, $P(\Sigma,k)\propto k^\psigmaex$, in Figure
\ref{fig.PSigmaBoth}.  For the range $\kkmin\in[2,30]$ at early times, we find
$\psigmaex = -1.39,-1.61,$ and $-1.53$ for $\betal,2$, and $20$, respectively,
roughly consistent with the addition of $-1$ to $\prex$ from the integration.
\citet{Padoan04c} measured $P(\Sigma,k)$, 
for three star-forming clouds, Perseus, Taurus, and the Rosetta nebula.  They
found significantly steeper spectra than we do here, $\approx-2.8$ for all three
clouds.  They also
performed synthetic observations through two simulations with $\mach=10$, a \sa\
simulation and an equipartition model, and find $\psigmaex=-2.25$ for the
equipartition model and $-2.7$ for the \sa\ model.  The difference in slope
\hlthree{between the slopes they found and the slopes in our simulation is}
most likely due to the observations and radiative transfer models missing the
high density material due to the limited dynamic range, as $^{13}\rm{CO}$
freezes onto dust grains at densities above $10^4 \percc$ \citep{Bacmann02}.  As discussed in \citet{Beresnyak05} and shown in Figure
\ref{fig.PlnRhoAll}, the slope of the density power spectra are quite sensitive
to the rare high density material, \hlthree{so even a slight decrease in the dynamic
range of the observations will cause a steepening in the spectra}.  To properly
compare, we will need to perform similar synthetic observations of our
simulations.  



\subsection{Density Power Spectra in the Collapsing State}\label{sec.PrhoCollapse}
\begin{figure} \begin{center}
\ifpdf
\includegraphics[width=\hw\textwidth]{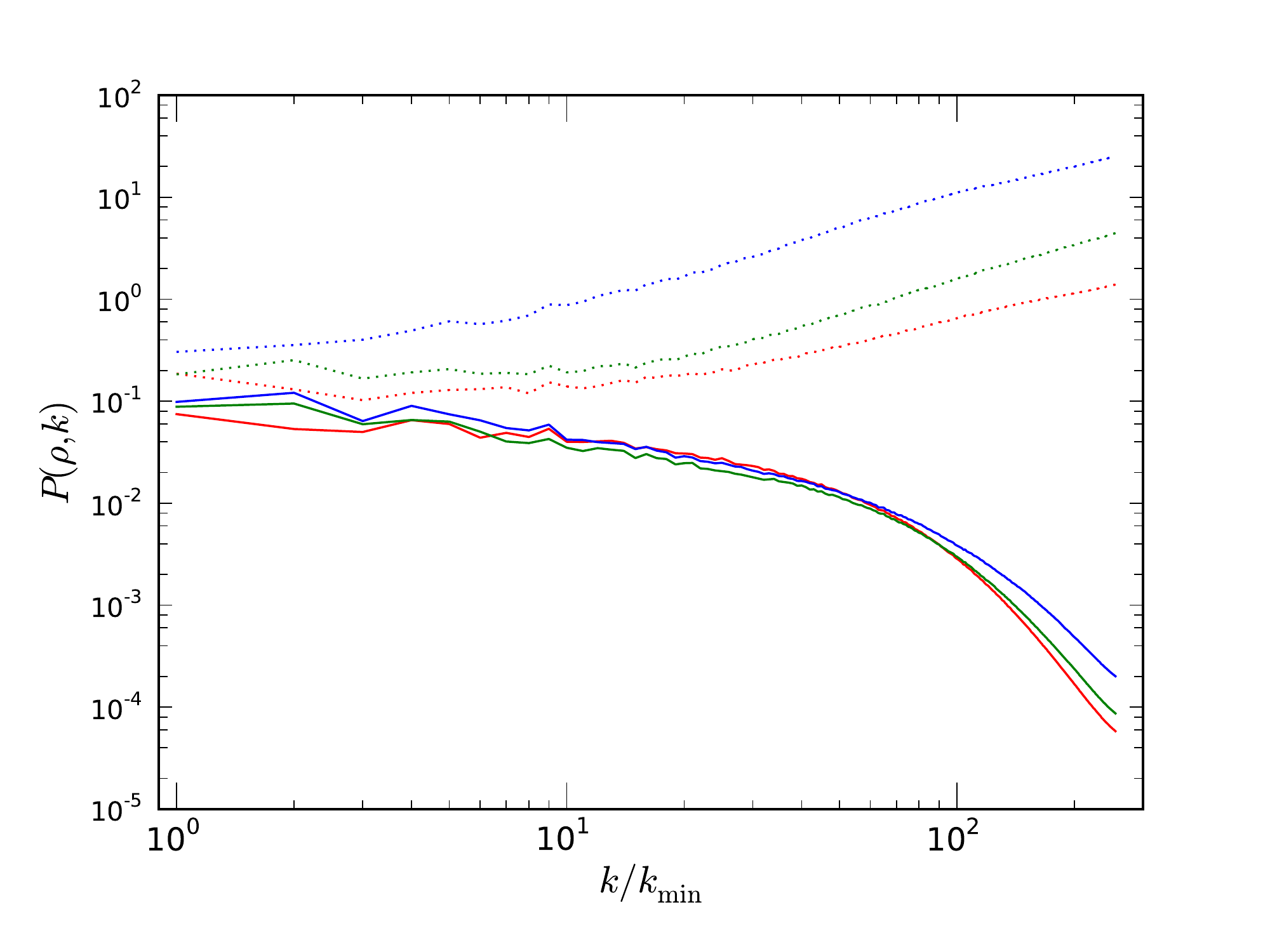}
\else
\includegraphics[width=\hw\textwidth]{figs_spectra//spectra_rhob025vb25vb205vDensity_4_6_8_10_57_59_61_63_65.eps}
\fi
\caption[ ]{Density power spectra $P(\rho,k)$ for all three simulations
(\betao=0.2, 2, 20 colored red, green, blue) and two snapshots, $t=0.1, 0.6\tff$
(solid, dotted).  \hlthree{The increasing nature in the later snapshot is expected from
self-similar spheres.}  }
\label{fig.PrhoBoth} \end{center} \end{figure}

The most prominent effect of gravity is the increase in slope of $P(\rho,k)$ with time, as seen
in Figure \ref{fig.PrhoBoth}.  For self-similar spheres with $\rho\propto
r^\rrex$, one finds $\prex=-2(\rrex+1)$.  Thus, positive slopes will be seen
in $P(\rho,k)$ for any $\rrex<-1$. Expected values from different self-similar
models are shown in Table \ref{table.self_similar_exponents}.  Our measured
values of $\prex$ for $\kkmin\in[10,200]$ are $0.86, 1.12, 1.2$ for \betao=0.2,
2, 20, respectively, at \tlate.  Direct comparison between the measured value of
$\prex$ and those predicted from the self-similar collapse models should be
handled carefully for two reasons, \hlthree{both stemming from to the volume weighted
nature of power spectra}:  first, he measured value contains
contributions from both the collapsing state and the turbulent state, which will
tend to decrease $\prex$ from the values expected from pure self-similar
spheres; second, the power spectrum contains contributions from unresolved gas
at very high densities.

The effect of increasing mean magnetic field on $P(\rho,k)$
in the collapsing state is to decrease the amount of power at all scales,
increase the wavenumber at which the slope becomes positive, and slightly
decreasing the slope.  This is consistent with the increased compressibility and
increased rate of collapse found in the more weakly magnetized simulations.  

The
column density power spectral slope, $\psigmaex$, becomes nearly flat for the
collapsing case, \hlthree{with $\psigmaex = -0.11, 0.15,$ and $0.25$ for
\betarespectively.}  This differs greatly from what is observed, owing to the fact
that the material causing the flat spectra is extremely high density, with
$\rho>(10^4-10^8) \rho_0 \approx (10^6-10^{10}) \percc$, wherein typical observational
tracers such as CO are frozen onto grains, and an extremely small volume
filling fraction, requiring high resolution observations.  High density tracers,
such as $\rm{NH}_3$ or deuterated species such as $\rm{H}_2 D^+$, will be
necessary to observe this signal \citep{Walmsley04,diFrancesco07}.  Synthetic
observations of our data, as well as high resolution observations in a broad
range of chemical tracers, will be required to further reconcile this
discrepancy.  Again, increased power with increased \betao\ is consistent with
increased compressibility of the gas.

\section{Energy Ratios}\label{sec.Energy}
\begin{figure} \begin{center}
\ifpdf
\includegraphics[width=\hw\textwidth]{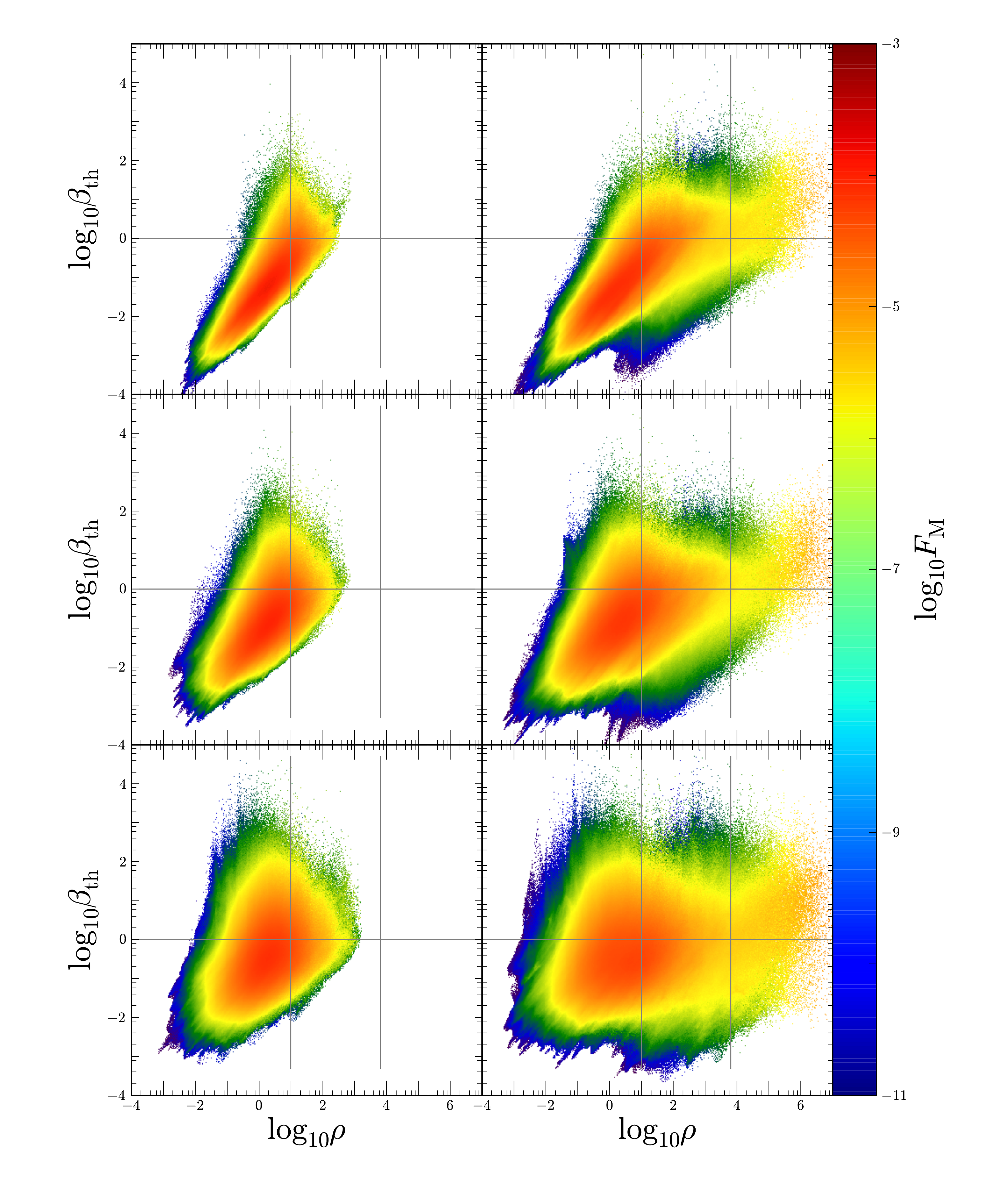}
\else
\includegraphics[width=\hw\textwidth]{figs_fielddensity/Betan_b025v_0010_0063b25v_0010_0063b205v_0010_0063_Phase_OverDensity_FieldNorm_CellMass.eps}
\fi
\caption[ ]{Thermal-to-magnetic pressure ratio, $\betath$, vs. $\rho$, colored by
mass fraction, $F_{\rm{M}}$, at $t=\tearly$ (left column) and $t=0.6 \tff$ (right columnd).  Top to bottom,
\betal,$0.2$ and $20$.  Increased mean field increases the correlation between
$\beta$ and $\rho$ and increasing the slope.  High density gas has largely
$\betath\approx 1$, showing a tendency towards pressure balance in collapsing
gas.}
\label{fig.Ptherm} \end{center} \end{figure}

The balance of energies sheds important light on the physical processes at play
in these clouds.  Figure \ref{fig.Ptherm} shows thermal-to-magnetic pressure
ratio, $\betath=8 \pi \cs^2 \rho/B^2$, versus density $\rho$; and Figure
\ref{fig.Pdyn} shows dynamic-to-magnetic pressure ratio, $\betadyn=8 \pi \rho
v^2/B^2$.  In both figures, the left column is taken at \tearly, the right
from \tlate, and top to bottom show $\betal,\ 2$, and 20, respectively.  Both
figures are colored by mass fraction $F_{\rm{M}}$.  

\subsection{Energy Ratios in the Turbulent State}
\begin{figure} \begin{center}
\ifpdf
\includegraphics[width=\hw\textwidth]{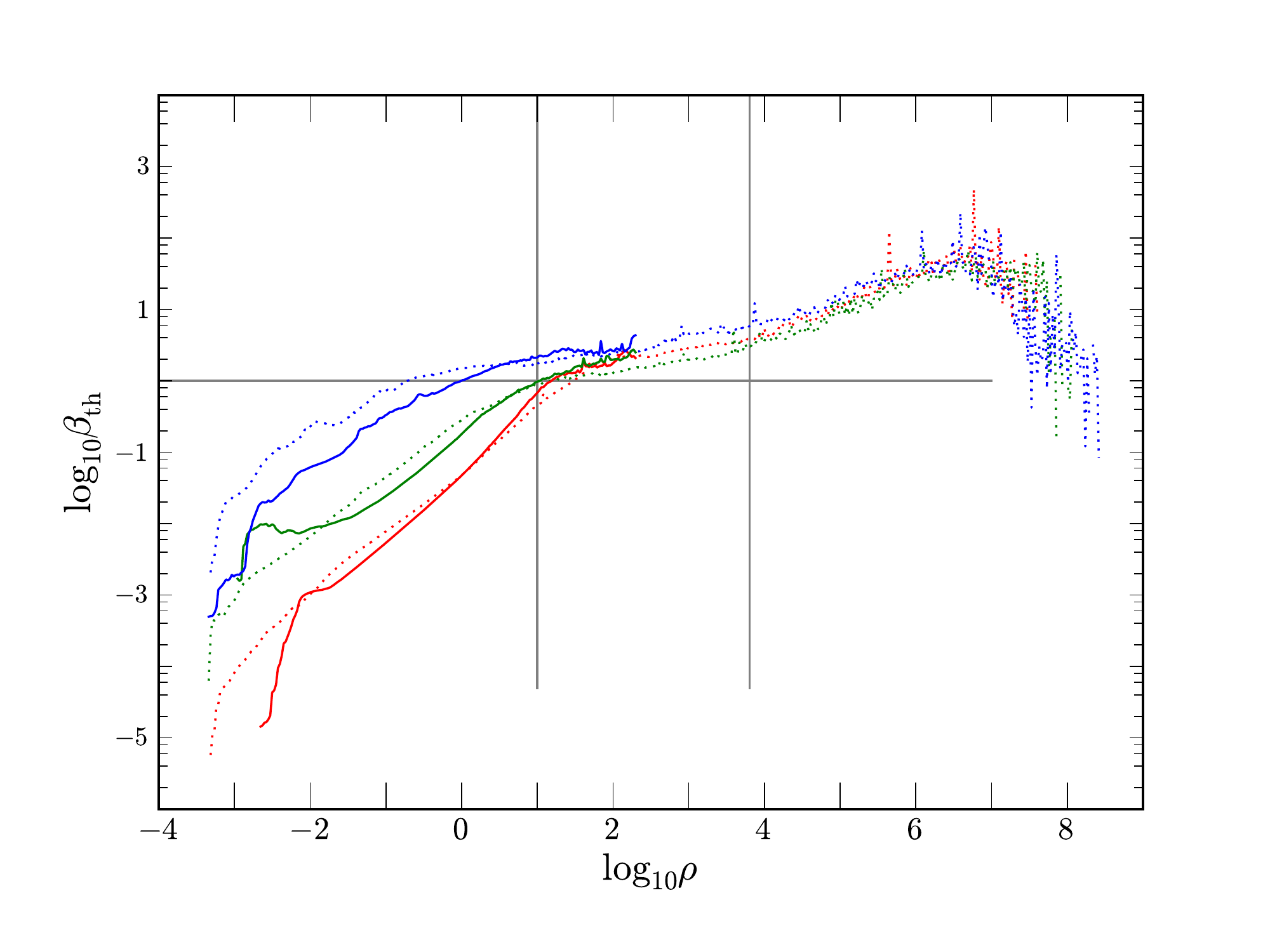}
\else
\includegraphics[width=\hw\textwidth]{figs_fielddensity/betarho_all3_low_high.eps}
\fi
\caption[ ]{Average thermal-to-magnetic pressure ratio $\betath$ vs. $\rho$
for all three simulations, \betao=0.2, 2, and 20 shown in red, green, and blue
respectively, at two snapshots, \tearly\ (solid) and \tlate\ (dotted.)  The
change in behavior, from turbulent to collapsing, is apparent at
$\betath\approx1$.  The horizontal line shows $\betath=1$, and vertical lines
show the three density regimes, as in Figure \ref{fig.VrhoBoth}.  This
highlights the fact that density alone is not enough to describe the transition
from turbulent to collapsing gas.}
\label{fig.BetaRho1d} \end{center} \end{figure}

In the turbulent state, $\betath$ shows some interesting variation with mean
field.  The scatter in $\betath$ increases with increasing \betao, and the mean
slopes in the $\betath\propto \rho^\betarex$ decreasing with increasing \betao;
$\betarex=0.96,\ 0.77,$ and $0.54$ for $\betal,\  2,$ and $20$, respectively, for
$\rho\in[10^{-2},10]$.  The average of $\betath$ versus $\rho$ can be seen in Figure
\ref{fig.BetaRho1d}.  For perfectly spherical
contractions, $B\propto\rho^{2/3}$, since $B\propto R^{-2}$ due to flux
conservation, and $\rho\propto R^{-3}$ due to mass conservation.  For flow
perfectly aligned with the field, $B\propto \rho^0$, since no amplification of
the field can take place, and for flow completely perpendicular to the field,
$B\propto\rho^1$ \citep[eg.][]{Kulsrud04}.  Since $\betath=\cs^2 \rho/B^2$, and for
an isothermal equation of state $\cs$ is constant, one finds
$\betarex= -1,\ -1/3,$ and $1$ for  perpendicular, isotropic, and parallel flow,
respectively.  \hlthree{This indicates that flow is preferentially aligned in all states,
even in the weakest field simulation, but the alignment increases with mean
field strength}  

\begin{figure} \begin{center}
\ifpdf
\includegraphics[width=\hw\textwidth]{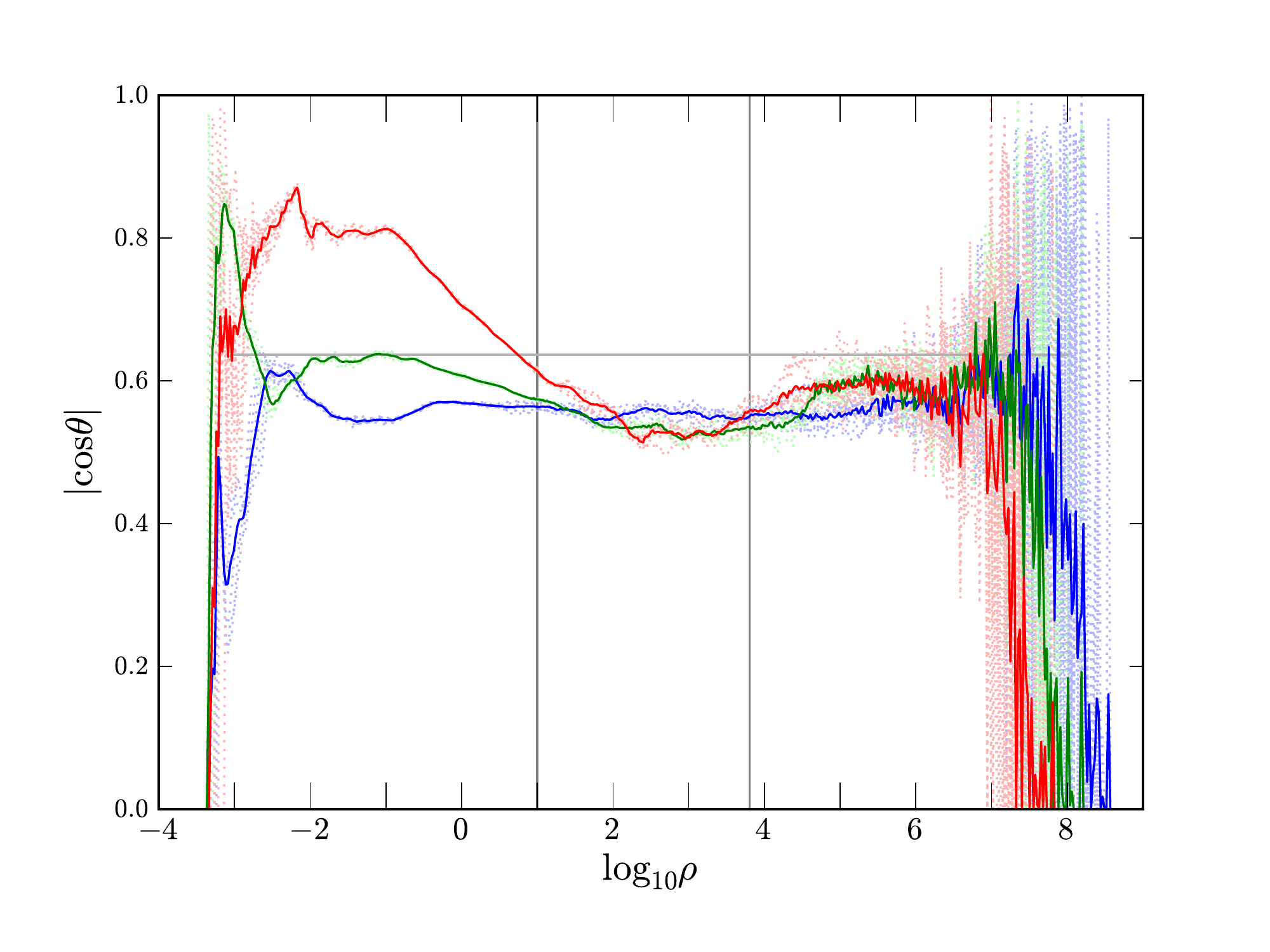}
\else
\includegraphics[width=\hw\textwidth]{figs_alignment//BVangleb205vb25vb025vBVangle_55_57_59_61_63_65.eps}
\fi
\caption[ ]{Angle between \bvec\ and \vvec\ ($|\cos\theta|$) for a number of late-time snapshots
($t=0.57, 0.59, 0.61, 0.63, 0.65\tff$, light dotted lines) and the average (red, green,
and blue for \betal, \betam, \betah, respectively). The horizontal line shows
the expectation for random $\theta$, while the two vertical lines separate
turbulent (left), collapsing (center) and unresolved (right) states.}
\label{fig.BVangle} \end{center} \end{figure}

The degree to which the velocity and field are aligned is shown
in Figure \ref{fig.BVangle}, which shows the mass weighted average of the magnitude of the
cosine of the angle between $\bvec$ and $\vvec$,
\def\ii{\rm{i}}
\begin{align}
\langle |\cos \theta| \rangle_{\rho_{\ii}} = 
\left\langle \frac{|\bvec\cdot\vvec|}{B v} \right\rangle_{\rho_{\ii}},
\end{align}
where the average is done over only material with density in bin $\rho_{\ii}$
and normalized to the total mass in that bin, and $512$ bins were used.
The grey line shows the expectation value of $|\cos \theta|$ for uncorrelated
vectors, $3/\pi\approx0.64$.  The solid red, green, and blue lines show $\betal, 2$,
and $20$, respectively, averaged for several snapshots around $\tlate.$  Light
dotted lines show the constituent snapshots, which demonstrates the extremely
short timescale on which $\theta$ varies at high density.  The $\betal$ simulation has
$\bvec$
and $\vvec$ nearly aligned at low density, which is consistent with
$\betarex=0.96\approx 1$, while the
other weaker field simulations show less alignment, even a slight tendency for $\bvec$
and $\vvec$ to be perpendicular, and correspondingly lower
values and larger variance in $\betarex$. 

\begin{figure} \begin{center}
\ifpdf
\includegraphics[width=\hw\textwidth]{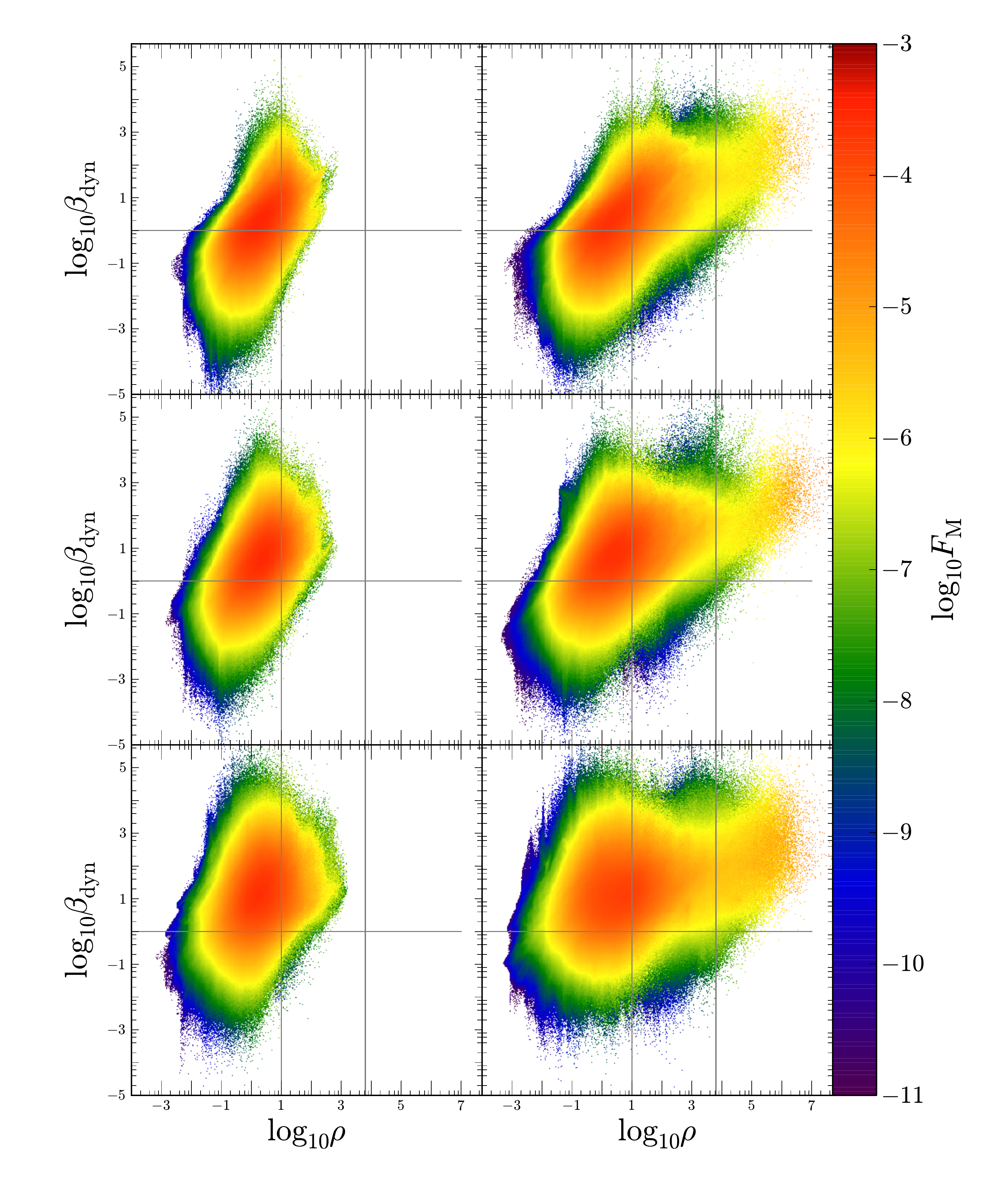}
\else
\includegraphics[width=\hw\textwidth]{figs_state//KEBE2_b025v_0010_0061b25v_0010_0061b205v_0010_0061_KEtoBEvsRho.eps}
\fi
\caption[ ]{Dynamic pressure ratio $\beta_{\rm{dyn}}$
for \tearly (left column) and \tlate (right column).  Top to bottom,
\betal,$0.2$ and $20$.   High density gas 
density gas is dominated by dynamic
pressure regardless of mean field strength.}
\label{fig.Pdyn} \end{center} \end{figure}

\hlthree{The left column of Figure \ref{fig.Pdyn} shows the ratio of
dynamic-to-magnetic pressure for the early snapshot.}
It can be seen from this figure that the typical gas
element in the low-density turbulent state in the $\betal$ run is trans-\alf, as $\betadyn=v^2/v_{\rm{A}}^2$ and
the gas  is evenly distributed around $\betadyn=1$, with a peak of the PDF at
$\betadyn=0.54$.  The other two simulations
are more \sa, with peak $\betadyn$ at 1.2 and 5.6.

\subsection{Energy Ratios in the Collapsing State}
For the high density collapsing gas, $\betath$ and $\betadyn$ can be seen in the
right columns of Figures \ref{fig.Ptherm} and \ref{fig.Pdyn}.   The transition
between turbulent and collapsing 
states is seen quite clearly in Figure \ref{fig.BetaRho1d}, \hlthree{which shows average
$\betath$ vs. $\rho$.  In this figure,} at 
$\betath\approx1$ the slope in mean $\betath$ flattens to almost zero, showing
that on average, collapsing gas is in thermal-to-magnetic pressure balance.  For
$\rho\in[100,\rhohigh]$, $\betarex = 0.15, 0.19,$ and $0.23$ for $\betal, 2$,
and $20$, respectively.  A lower limit of $\rho=100$ was used for these fits as
this is approximately where the slope in the $\betal$ simulation flattens
dramatically.   However, the density at which this transition in slope happens
is found to decrease
with increasing magnetic field, with 
 $\rho\approx 10$ and $0.1$ for $\betam$ and $20$,
respectively.  This indicates that density alone is not a sufficient variable to
mark the transition from turbulent to collapsing gas.  

As shown in the right column of Figure \ref{fig.Pdyn}, the collapsing gas is
dominated by dynamic pressure for all three values of $\betao$.  This is also
true for the unresolved gas, which while one cannot quantitatively trust these
results as they are contaminated by numerical resolution problems, the
likelihood  that increased resolution would decrease this ratio by 
two orders of magnitude is low.  This
suggests that collapsing state is formed from gas that is initially \sa, where
magnetic energy support is insufficient to resist ram pressure from the gas,
causing density peaks that can become gravitationally bound.  As the mass
fraction of gas that is \sa\ decreases with \betao\, this also has the effect of
decreasing the amount of gas that can collapse to high densities, in turn
decreasing the rate and efficiency of star formation.  In non-self-gravitating
turbulence results, the existence of \sa\ high-density gas is seen even in sub-sonic,
sub-\alf\ simulations \citep{Burkhart09}, as well as in thermally unstable trans-\alf\
simulations \citep{Kritsuk11c}.  Gravity, however, has the effect
of highly concentrating the high density gas in the \sa\ regime.  
  This will be compared with observations in Section \ref{sec.discussion}.


\section{Magnetic Field PDF}\label{sec.VB} 
Figure \ref{fig.VB} shows PDF of the magnetic field for \tearly\ (left panel)
and both \tearly\ and \tlate\ (right panel).  The left plot is linear, and shows
$V(b)$, where $b$ is the fluctuating field, $b=|\bvec-B_0|$.   The
right plot is logarithmic, and shows the PDF for the full magnetic field
$V(B)$.

\subsection{Magnetic Field PDF in the Turbulent State}
\begin{figure*} \begin{center}
\ifpdf
\includegraphics[width=\hw\textwidth]{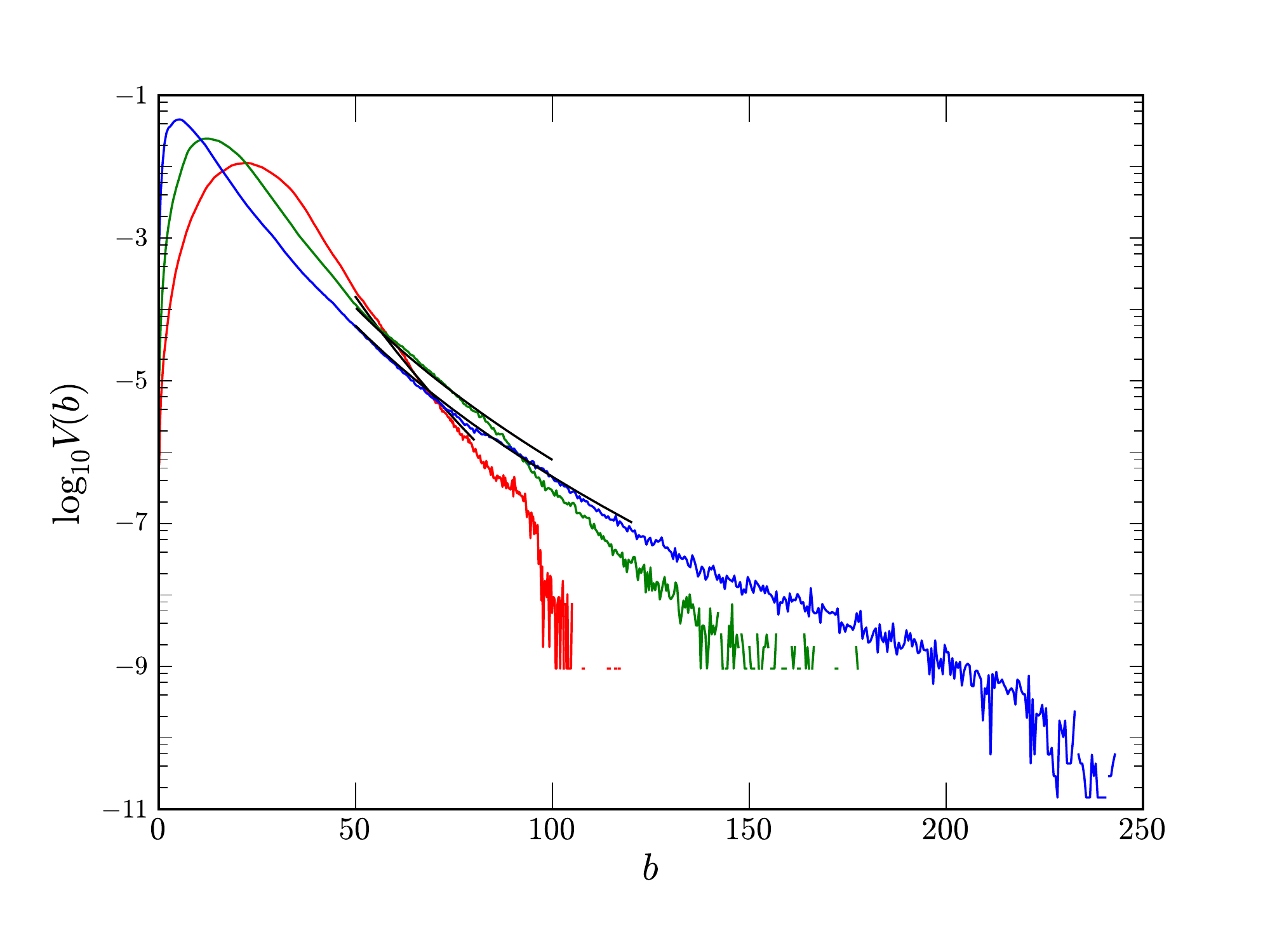}
\includegraphics[width=\hw\textwidth]{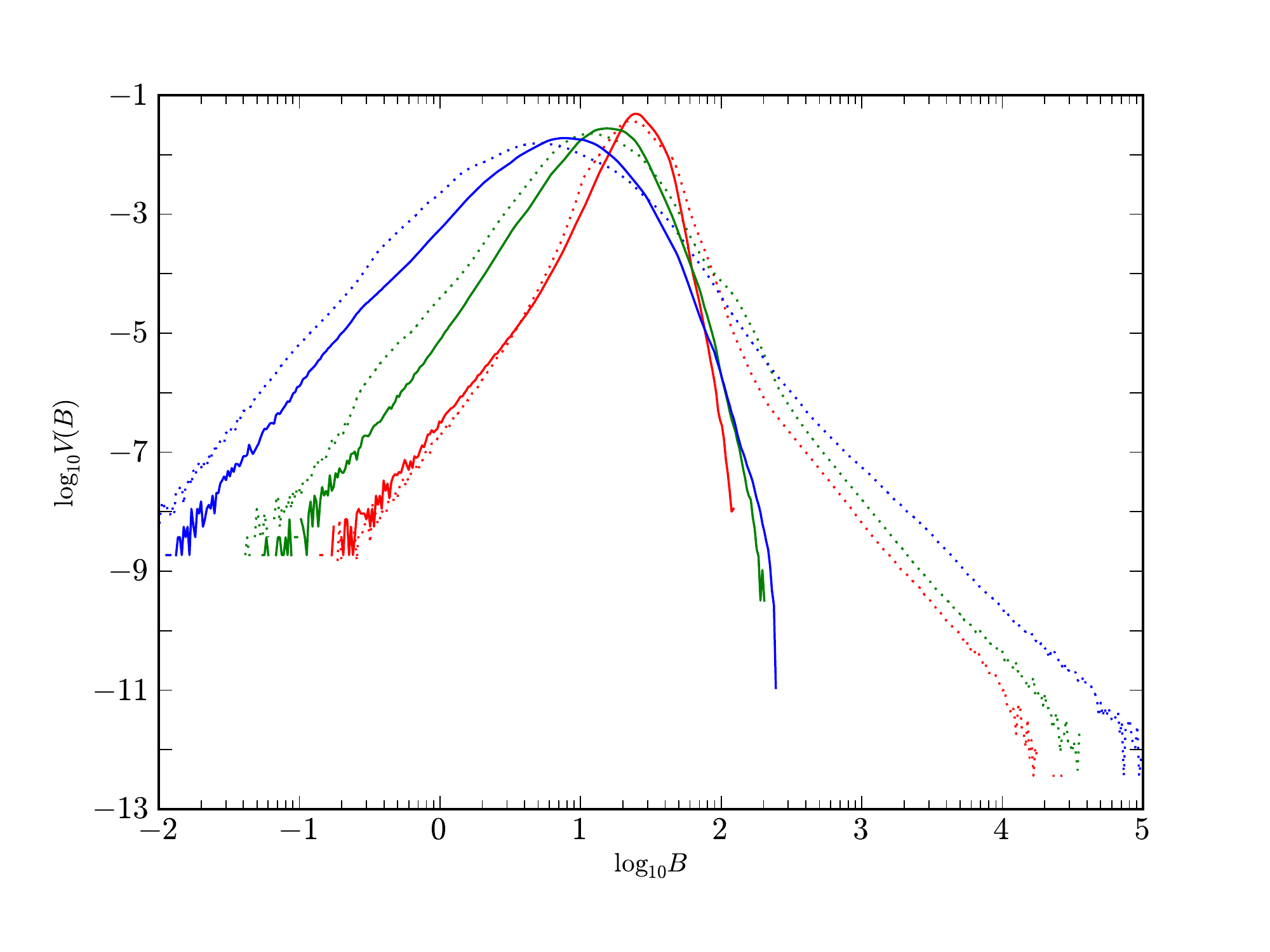}
\else
\includegraphics[width=\hw\textwidth]{figs_PDF//VBlinearb025vb25vb205vVolumeFraction_10.eps}
\includegraphics[width=\hw\textwidth]{figs_PDF//B_PDF_rhob025vb25vb205vVolumeFraction_4_6_8_10_57_59_61_63_65.eps}
\fi
\caption[ ]{PDF of magnetic field strength.  (Left) 
PDF of fluctuating $b=|\bvec-B_0|$, $V(b)$, for $\tearly$.  Fits to a
stretched exponential are shown in black for each line.  Note that this figure
is linear in $b$.  The smaller volume fractions achieved by the $\betah$
simulation are due to the more rapid collapse, thus earlier refinement, of that
simulation.  (Right) PDF of the full magnetic field strength, $B$, with
lines colored the same as in Figure \ref{fig.VrhoBoth}.  Here we see a power-law
developed at late times due to the collapsing gas.  This figure illustrates how
turbulence and gravity leave different signatures in the PDF of magnetic field
strength; at high field strength, a stretched-exponential tail is generated by turbulence, while a
power-law tail is generated by gravity.}
\label{fig.VB} \end{center} \end{figure*}

\begin{figure} \begin{center}
\ifpdf
\includegraphics[width=\hw\textwidth]{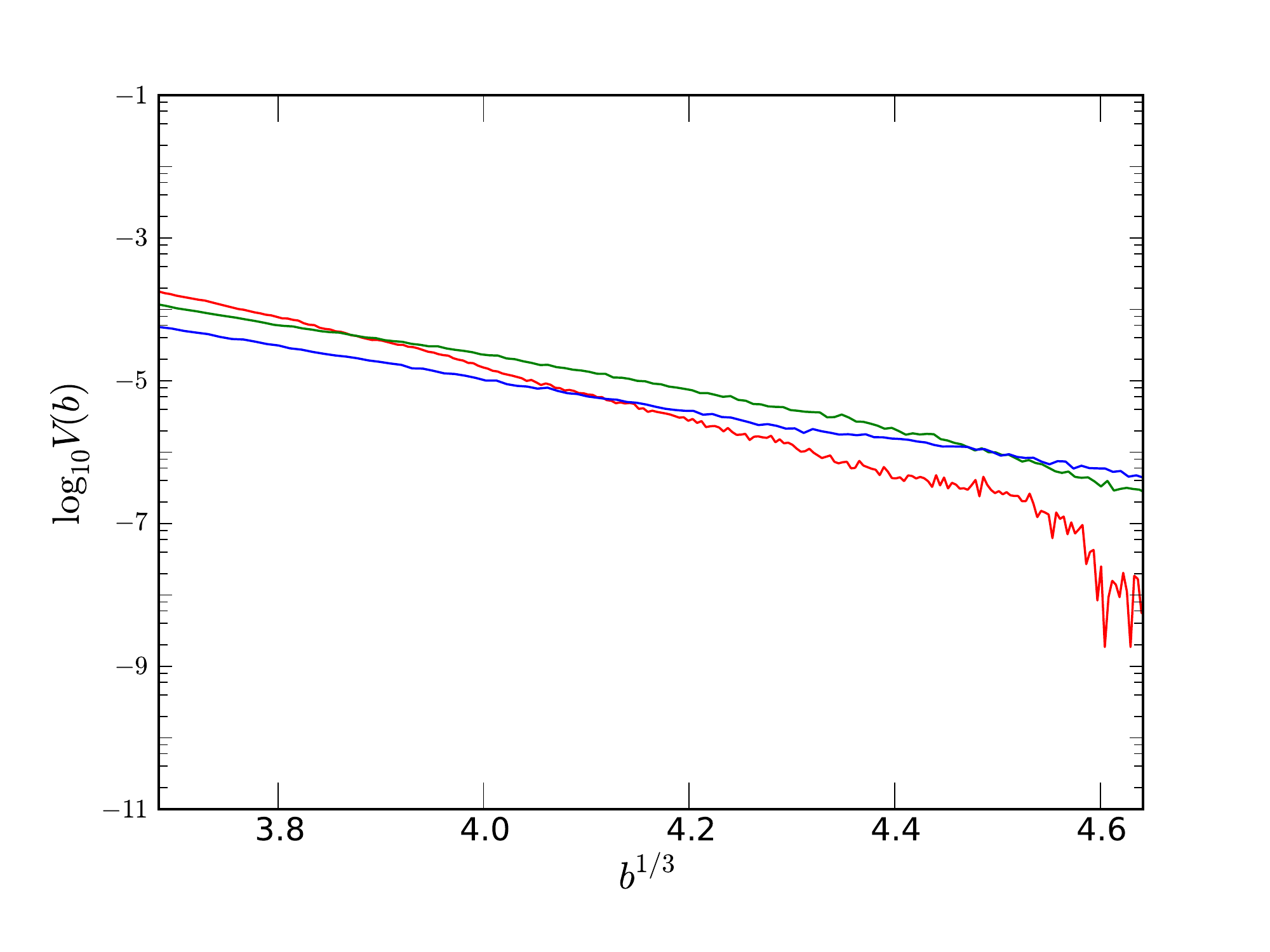}
\else
\includegraphics[width=\hw\textwidth]{figs_PDF//VBonethirdb025vb25vb205vVolumeFraction_10.eps}
\fi
\caption[ ]{The stretched exponential section of $V(b)$ at $\tearly$, restricted
to the fit range $b\in[50,100]$, and plotted against $b^{1/3}$.}
\label{fig.VBonethird} \end{center} \end{figure}

Kritsuk et al. (2012, in preparation)
have found that the high field wing of the PDF of magnetic
field in an isothermal turbulent
gas can be well described by a stretched exponential, of the form 
\begin{align}
V(b) db = c(b^{c-1}/b_0^c) \exp \left [ -(b/b_0)^c \right] db,
\label{eqn.stretched}
\end{align}
with a stretching exponent $c\approx 1/3$.
A stretched exponential describes a sequence of multiplicative events, where
$1/c$ is the depth of the hierarchy of events \citep{Frisch97,Laherrere98}.
This is not unlike the sequence of multiplicative shocks that generates the
density PDF $V(\rho)$. 
For \betal\ we find $b_0=3.1\sci{-3}$ and $c=0.33$ for $b\in[50,80]$; for
\betam\ we find $b_0=6.1\sci{-3}$ and $c=0.32$, for $b\in[50,100]$; 
and for \betah, we find $b_0= 9.0\sci{-3}$ and $c=0.33$ for $b\in[50,120]$.
Fits to these lines can be seen in black along each curve in the left panel of
Figure \ref{fig.VB}
To better demonstrate the fit, Figure \ref{fig.VBonethird} shows $V(b)$
against $b^{1/3}$, restricted to the interval of the fit.  
Thus we find that the number of multiplicative events is the same for all three
simulations, $1/c\approx 3$, but the characteristic scale, $b_0$, increases
with $\betao$ due to the greater ease of inducing fluctuations in a weaker mean
field.  

\subsection{Magnetic Field PDF in the Collapsing State}
The collapsing state exhibits a power-law
$V(B)\propto B^\vbex$.  As an example, for the mid-strength field, \betam\, we
find $\vbex=-4$ for $B\in[52,530]$. This fit range was determined by the average
magnetic field spanned by densities in the range $\rho\in[\rholow,\rhohigh]$.
We compute the exponent $\brex$ in $B\propto\rho^{\brex}$, and find
$\brex=0.43$ for this simulation.  Combining this with 
$\vrex=-1.7$ found in Figure \ref{fig.vrex_convergence}, we predict
$\vbex=-3.9$, in reasonable agreement with the measured value.  

The slope of $V(B)$ increases with increasing mean field strength.
For the \betah\ case, we find a fit exponent $\vbex=-3.22$ for $B\in[27,670]$,
while for \betal\ we find $\vbex=-5.42$ for $B\in[36,451]$.   Using measured values
of $\brex$ and $\vrex$, we predict slopes for
$V(B)=-4.5, -4.0,$ and $-3.3$ for $\betal$, 2, and 20, respectively.  The
predicted value of $\vbex$ for the  $\betal$ run is somewhat lower than the
measured value.  This is likely due to the fact that the run had not yet fully
developed the collapsing state.

\section{Velocity and Energy Power Spectra}\label{sec.Pv}
The velocity power spectrum, $P(v,k)$, is historically one of the most studied 
quantities in turbulence modeling.  Figures \ref{fig.PvFull} and
\ref{fig.PvBoth} show compensated velocity spectra.  One curious feature is that
the collapsing state doesn't show up in the velocity spectra, and only shows a mild
feature in the kinetic energy spectra, $P(\rho^{1/2} v,k)$, and magnetic energy
spectra, $P(B,k)$, as seen in Figure \ref{fig.PengBoth}.

\subsection{Velocity and Energy Power Spectra in the Turbulent State}
\begin{figure} \begin{center}
\ifpdf
\includegraphics[width=\hw\textwidth]{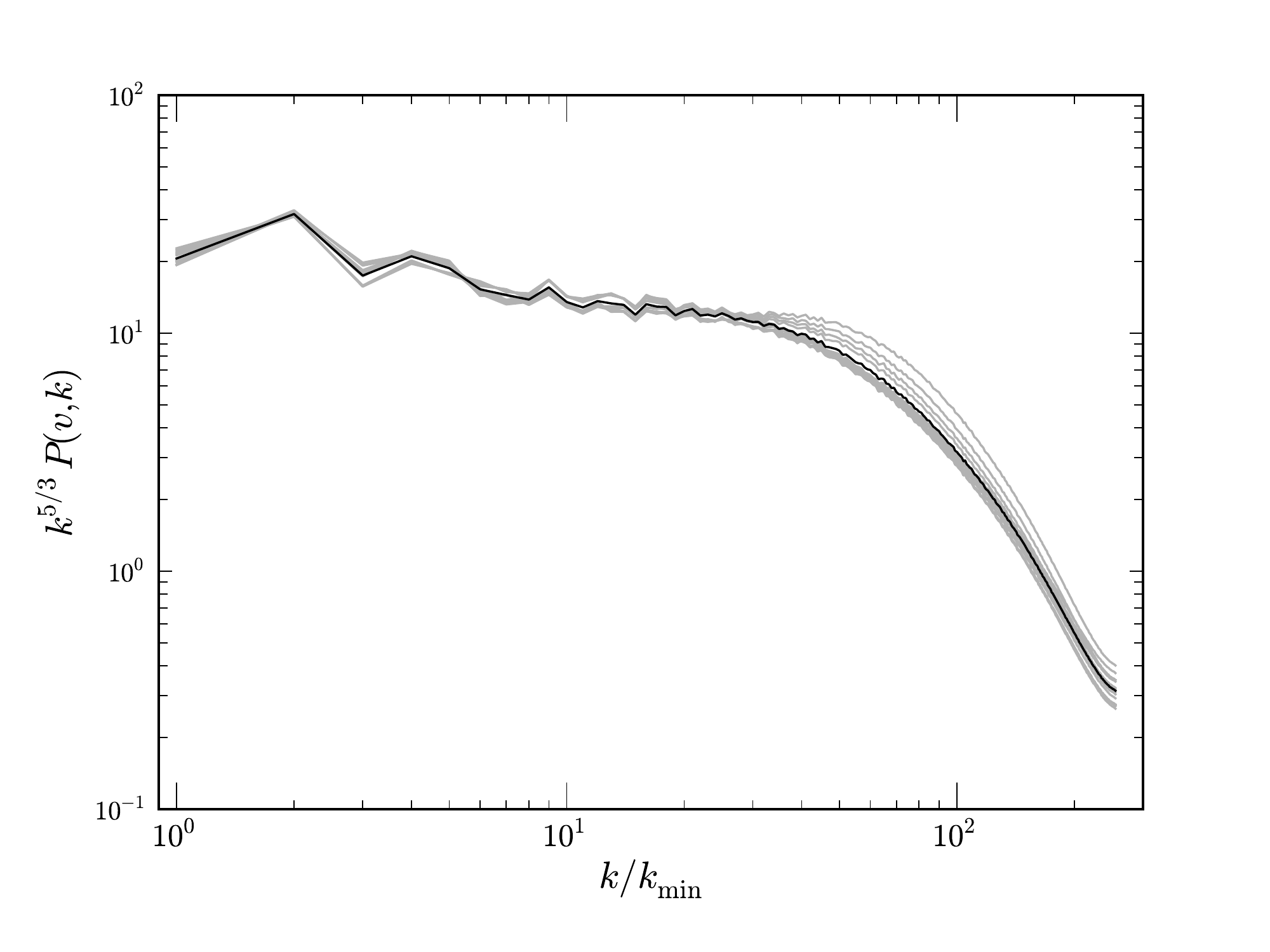}
\else
\includegraphics[width=\hw\textwidth]{figs_spectra//spectraV0b205vvelocity_4_6_8_10_25_27_29_31_33_55_57_59_61_63_65.eps}
\fi
\caption[ ]{Compensated velocity power spectrum $k^{5/3} P(v,k)$ for 15 snapshots
between $t=0.0-0.65 \tff$ for the \betah\ simulation.  All the evolution seen
can be attributed to relaxation from the initial conditions, no evolution due to
gravity is seen.}
\label{fig.PvFull} \end{center} \end{figure}

Only the turbulent state is visible in the velocity power spectrum, $P(v,k)$.  
Figure \ref{fig.PvFull} shows the compensated
velocity power spectra, $k^{5/3} P(v,k)$, for all snapshots for the \betah\
simulation (grey lines) and the average of all snapshots (black line).  The only
variation comes from the first $0.1\tff$, where the variation is due entirely to
the change in the effective spectral bandwidth between PPML, the solver used for
the initial conditions, and the solver in Enzo, which was used for the self-
gravitating AMR simulations.  The rest of the snapshots, nearly
indistinguishable from the mean, show almost no evolution
at all.   The other two simulations (not shown) show even less
variation in the initial phase.

\begin{figure} \begin{center}
\ifpdf
\includegraphics[width=\hw\textwidth]{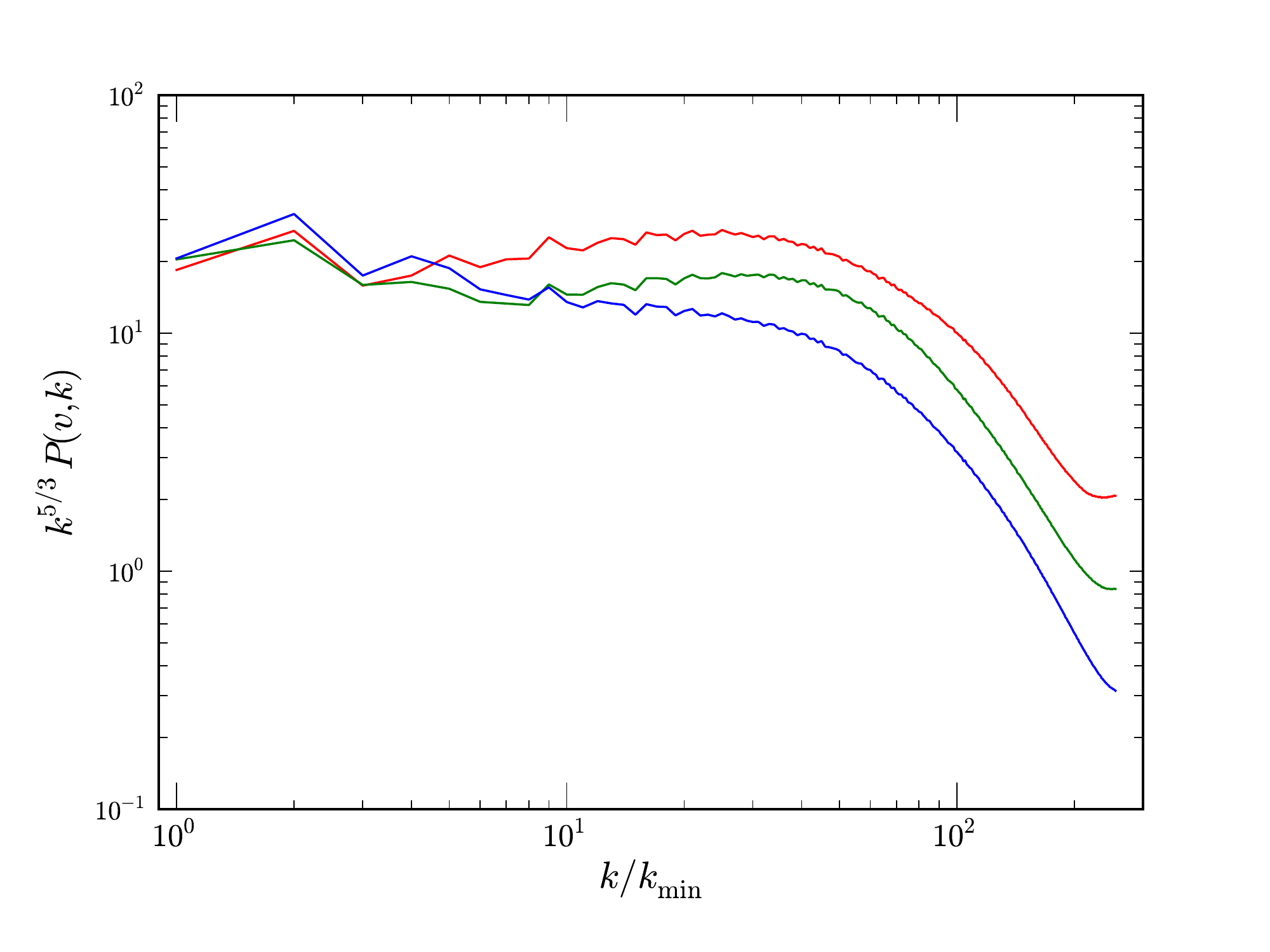}
\else
\includegraphics[width=\hw\textwidth]{figs_spectra//spectraV0b025vb25vb205vvelocity_4_6_8_10_25_27_29_31_33_55_57_59_61_63_65.eps}
\fi
\caption[ ]{Velocity power spectra $P(v,k)$ for all three simulations, averaged
over all times.  Slopes for $\kkmin\in[2,30]$ are $-1.46$, $-1.58$, and $-1.80$
for $\betao=0.2, 2, 20$,
respectively.
}
\label{fig.PvBoth} \end{center} \end{figure}

Figure \ref{fig.PvBoth} shows compensated power spectra for all three
simulations, with red, green, and blue showing  \betal, 2, and 20, 
respectively.  Averages were taken over all snapshots, as the variation for
\betam\ and \betah\ simulations were even smaller than that shown in Figure
\ref{fig.PvFull}.  Clearly, the mean field strength has a significant effect on
the spectral slope.  
\hlthree{While the resolution of these simulations is too low to make precise
measurements of the slope in the inertial range, measuring the slope is useful
to compare scaling among the three simulations.}
Slopes for $\kkmin\in[2,30]$ are $-1.46$, $-1.58$, and $-1.80$ for $\betao=0.2,
\ 2,$ and  $20$,
respectively.  \hlthree{Though there is presently no theory to predict the slope
as a function of mean field in compressible MHD turbulence, the flattening of
the slope
with increasing mean field seen here is consistent with what one can infer from the
existing theories for compressible hydro and incompressible MHD.}  The \betah\ slope is consistent with other supersonic
hydrodynamic simulation of \citet{Kritsuk07}, and is the upper end of the slope predicted by
\citet{Boldyrev02}.  The slope for
the \betal\ simulation is reminiscent of the $-3/2$ value of the
Iroshnikov-Kraichnan model
\citep{Iroshnikov64,Kraichnan65}.
This reduction of slope with mean field strength was also
seen by \citet{Kritsuk09b}, who measure $-1.94,\ -1.62,$ and $-1.51$ for
$\betao=0.2,\ 2,$ and $20$, respectively.  This trend of increasing slope with
mean field was continued by
\citet{Lemaster09b}, who measured a slope of $-1.38$ for $\betao=0.02$,
which is flatter and of lower $\betao$ than our strong field run.  The \betam\ run
seems to be in transition between the two, \hlthree{with the slope increasing for
wavenumbers above} $\kkmin=10$.

The flattening of the spectra is consistent with the magnetic energy coming into
equipartition with the kinetic energy, as demonstrated by Figure
\ref{fig.PengBoth}.  The magnetic energy is about ten times lower than the
kinetic in the \betah\ simulation for all wavenumbers, so the similarity between this run and other
hydrodynamic runs is expected.  The \betam\ simulation displays equipartition
only in a very small band around $\kkmin=10$, and at higher wavenumbers, where
the magnetic energy is still below equipartition, the slope of
the velocity spectrum seems to
increase.  The \betal\ simulation has near
equipartition for the majority of the low to mid wavenumbers, and has the
flattest spectrum overall.

Another illuminating feature of Figure \ref{fig.PengBoth} is the
sub-equipartition nature of the magnetic field in these simulations at small
scales at early times (solid curves).  
This lack of equipartition at small scales is somewhat surprising, as standard
expectation of a small scale dynamo in incompressible MHD is to
first grow exponentially at small scales until equipartition is reached, then
linearly at larger scales, with the equipartition wavenumber decreasing with time \citep{Brandenburg05}.  However, as the
turbulence in this simulation is dominated by shocks, one cannot apply the same
physical arguments, since the shock jump conditions do not imply simple
equipartition between kinetic and magnetic energy in the shock-compressed layer.
The increased fraction of kinetic energy in compressive motions (see Section
\ref{sec.HelmholtzBoth}) generates less vorticity per unit of energy than
sub-sonic turbulence, which in turn generates less magnetic energy.  
This has been explored in \citet{Federrath11b}, who demonstrated that as Mach
number increases, the ratio of magnetic to kinetic energy transitions from near
unity for $\mach<1$, to a few percent for $\mach>1$.  This is consistent with
our findings here, though predominantly at small scales.  In our $\betam$ and
$\betal$ simulations, the imposed large scale field allows equipartition to be
reached at intermediate scales.

\subsection{Velocity and Energy Power Spectra in the Collapsing State}
As seen through the time evolution of $P(v,k)$ in Figure \ref{fig.PvFull}, the
collapse state does not leave a signature on the velocity power
spectrum.  This is due to the volume weighted nature of power spectra.  The
collapsing state occupies a very small volume fraction, as seen in Figure
\ref{fig.VrhoBoth}.  Because of this, for a signal to appear in the power
spectrum the values must be extremely large.  The values of velocity reached by
the gas do not vary by the many orders of magnitude that the density does in the
collapsing gas, as seen in Figures \ref{fig.VrhoBoth} and \ref{fig.PrhoBoth}.

If we relate the power spectra found here with the self-similar solutions
discussed in Section \ref{sec.VrhoCollapse}, we find that the pressure-free
collapse gives velocity scaling exponent $\pvex=+5/7$; the expansion wave
solution gives $\pvex=0$, which gives no visible signal to the velocity scaling;
finally the Larson-Penston solution, with constant velocity, gives no
contribution to the velocity spectrum.  

\begin{figure} \begin{center}
\ifpdf
\includegraphics[width=\hw\textwidth]{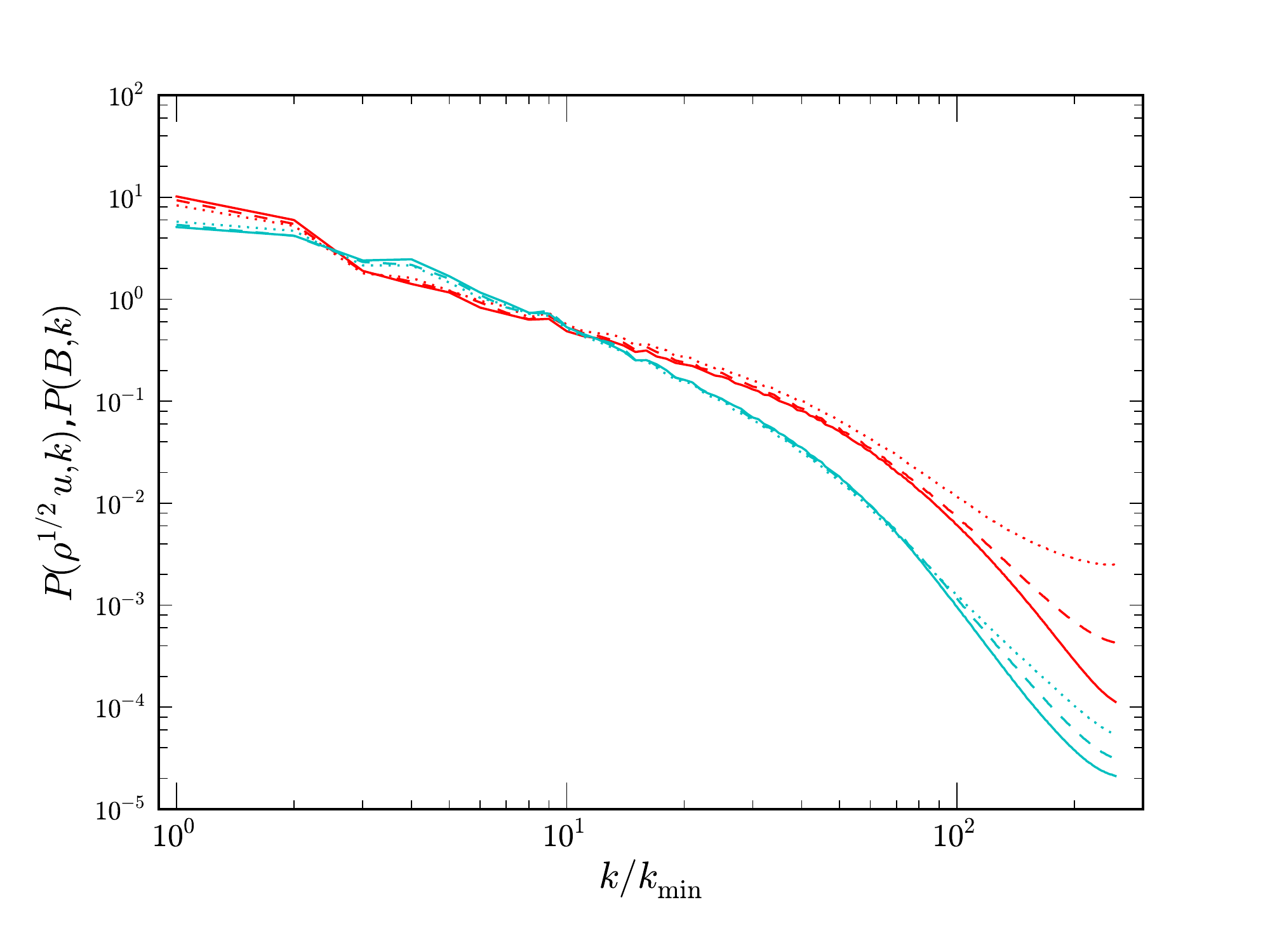}
\includegraphics[width=\hw\textwidth]{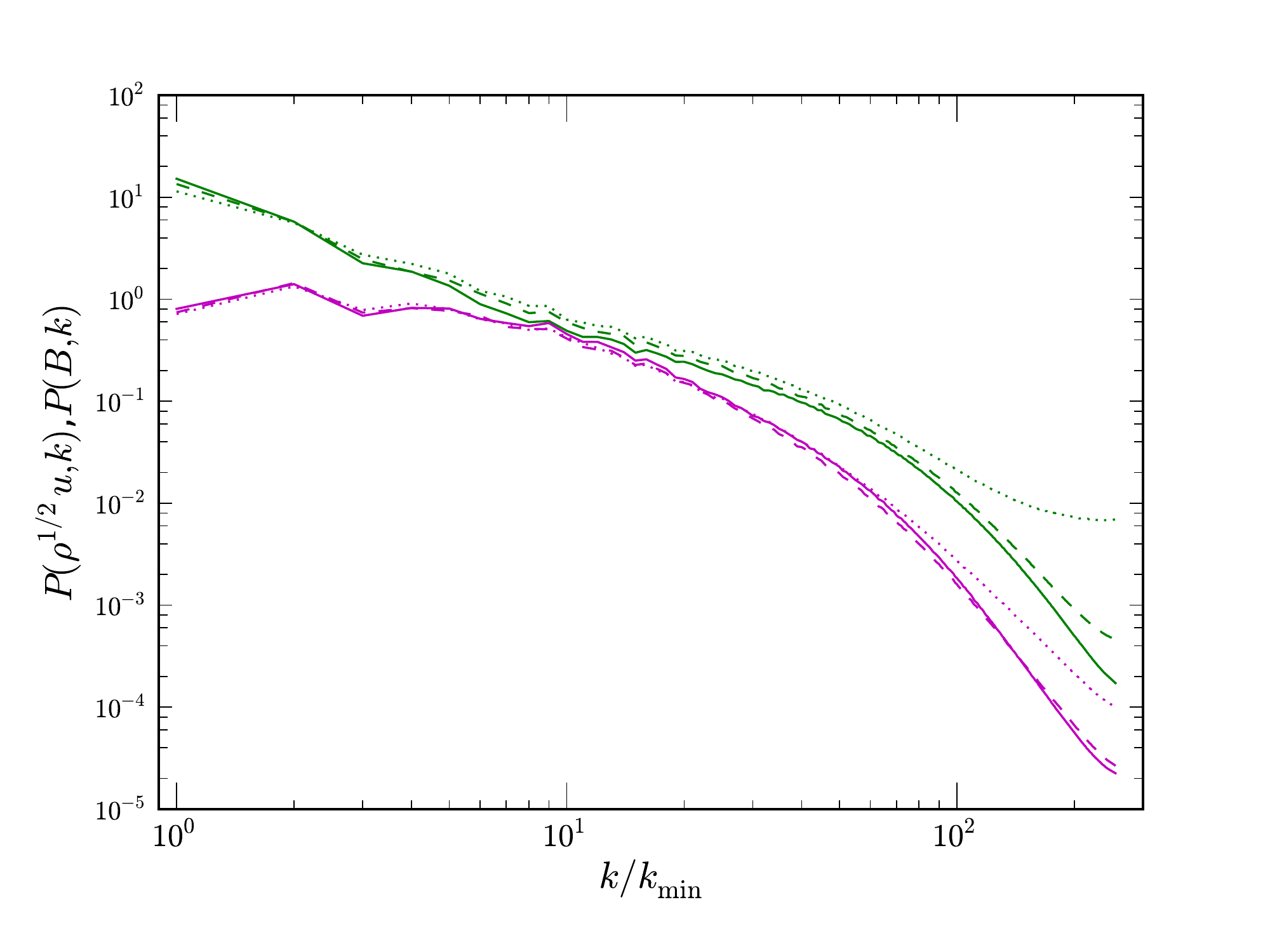}
\includegraphics[width=\hw\textwidth]{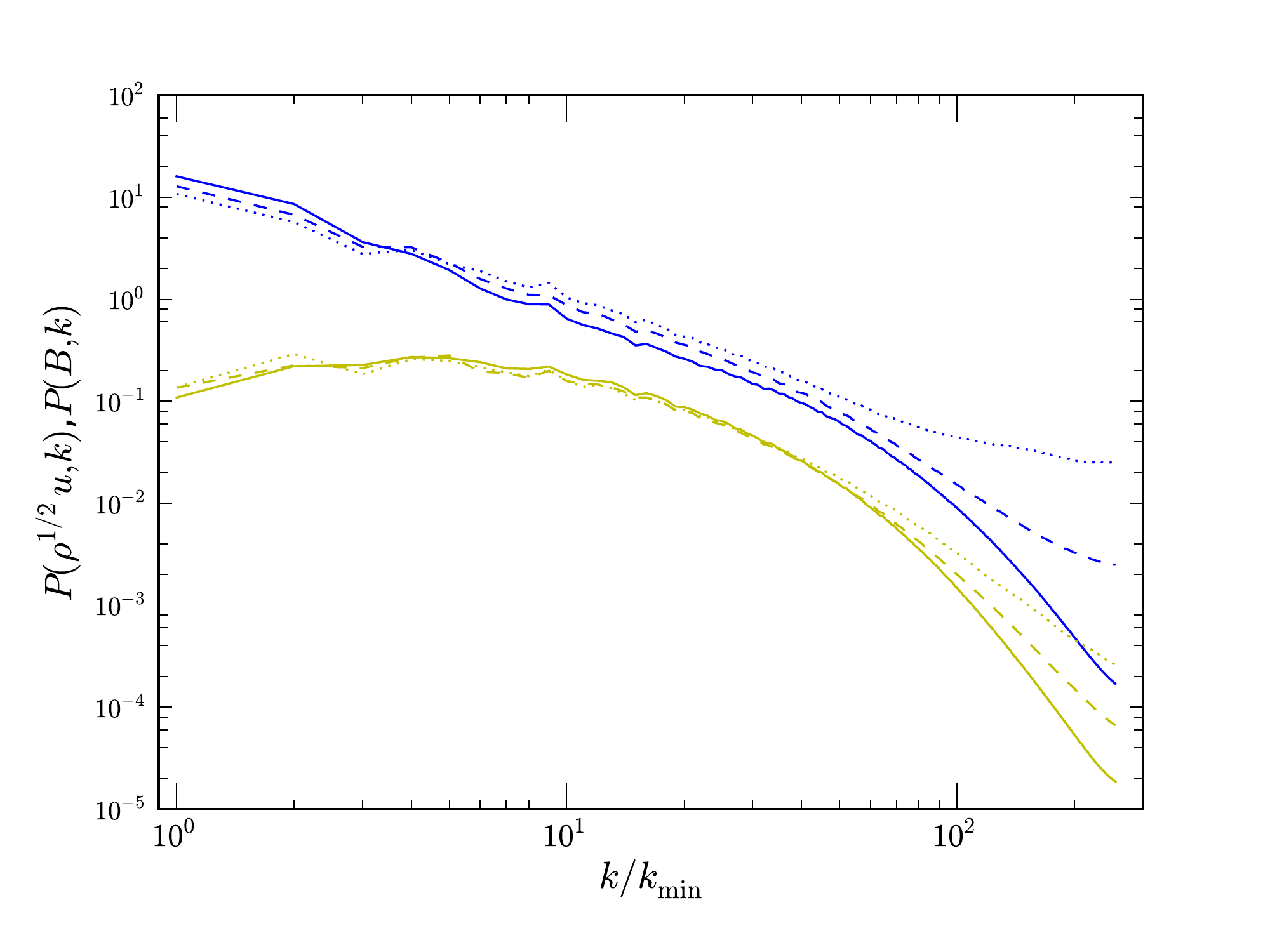}
\else
\includegraphics[width=\hw\textwidth]{figs_spectra//spectraV1_b025vMagnetic_10_35_61.eps}
\includegraphics[width=\hw\textwidth]{figs_spectra//spectraV1_b25vMagnetic_10_35_61.eps}
\includegraphics[width=\hw\textwidth]{figs_spectra//spectraV1_b205vMagnetic_10_35_61.eps}
\fi

\caption[ ]{Energy power spectra, $P(\rho^{1/2} v,k)$ (upper curves in each panel,
red, green, and blue)
and $P(B,k)$ (lower curves in each panel, cyan, magenta and yellow) for $\betal$
(top panel) $\betam$ (middle panel) and $\betah$ (bottom panel).  Three
different times are shown; $\tearly$ (solid line), $\tmid$ (dashed line), and
$\tlate$ (dotted line)}
\label{fig.PengBoth} \end{center} \end{figure}

While the collapsing state leaves no signature in the velocity spectra, it does
leave a signature at high $\kkmin$ on the kinetic and magnetic energy spectra,
$P(\rho^{1/2} v,k)$, and $P(B,k)$.  
In Figure \ref{fig.PengBoth} we show three snapshots for the energy spectrum,
$P(\rho^{1/2} v,k)$ (upper set of curves) and $P(B,k)$ (lower set), for all three
simulations (top to bottom, \betal, \betam, \betah).  Solid, dashed, and dotted
lines show $t=0.1, 0.3, 0.6\tff$, respectively. The kinetic spectrum shows an imprint
of the collapse in the high $\kkmin >100$ gas, due mostly to the density-weighted
nature of this statistic, and the fact that, on small scales, the density increases by as
much as 5 orders of magnitude.  Strictly speaking, the magnetic spectrum is also a volume weighted
quantity, so the increase in magnetic energy seen is a product of additional
field amplification from the collapse itself.  However, as discussed in Section
\ref{sec.Energy}, $B\propto \rho^{\brex}$, with $\brex\approx 0.4$, which is a
similar to the density dependence of kinetic energy, so $P(B,k)$ can be
considered to be implicitly density weighted.  The initial (turbulent) field
distribution is a product of the shock dominated turbulence, as discussed in
the previous section.  After gravity is introduced, the field sees
additional amplification due to the collapse itself.  This field amplification
is again consistent with increased compressibility due to increased $\betao$, as
the increase in $P(B)$ at high $\kkmin$ is greater for the weaker field runs.

If one
examines the energy scaling in the context of self-similar spheres, one finds that $P(\rho^{1/2} u)$ scales as $k^{+1}$
for the Larson-Penston  and pressure-free solutions, and $k^{+3/7}$ for the
expansion wave solutions.  These increasing solutions are not seen in Figure
\ref{fig.PengBoth}.  Again this is likely due to the fact that the superposition of the
turbulent and collapsing spectra makes distinguishing between the two
difficult.

\section{Helmholtz Decomposisition }\label{sec.HelmholtzBoth}
The formation of high density material necessarily comes from compressive
motions.  
A standard
method of examining the compressional and solenoidal modes of the velocity field
is the Helmholtz decomposition.  We split the velocity field, \vvec, into
two components
\begin{align}
\vvec = \vc& + \vs,\\
\nabla\cdot\vs &= 0,\\
\nabla\times\vc &= 0,
\end{align}
where we find \vc\ and \vs\ in Fourier space,
\begin{align}
\tilde{\vvec}_{\rm{c}} &= (\hat{k}\cdot\Fourier{\vvec})\hat{k},\\
\tilde{\vvec}_{\rm{s}} &= \Fourier{\vvec} - \tilde{\vvec}_{\rm{c}}
\end{align}
where \Fourier{\vvec} is the Fourier component of \vvec, and $\hat{k}$
is the unit wave vector.  We then examine the ratio of the power
spectra $\chi(k) = P({\vc},k)/P({\vs},k)$, which gives us a scale-wise measure of the importance
of compressibility of the gas.  

\subsection{Helmholtz Decomposition in the Turbulent State}
\begin{figure} \begin{center}
\ifpdf
\includegraphics[width=\hw\textwidth]{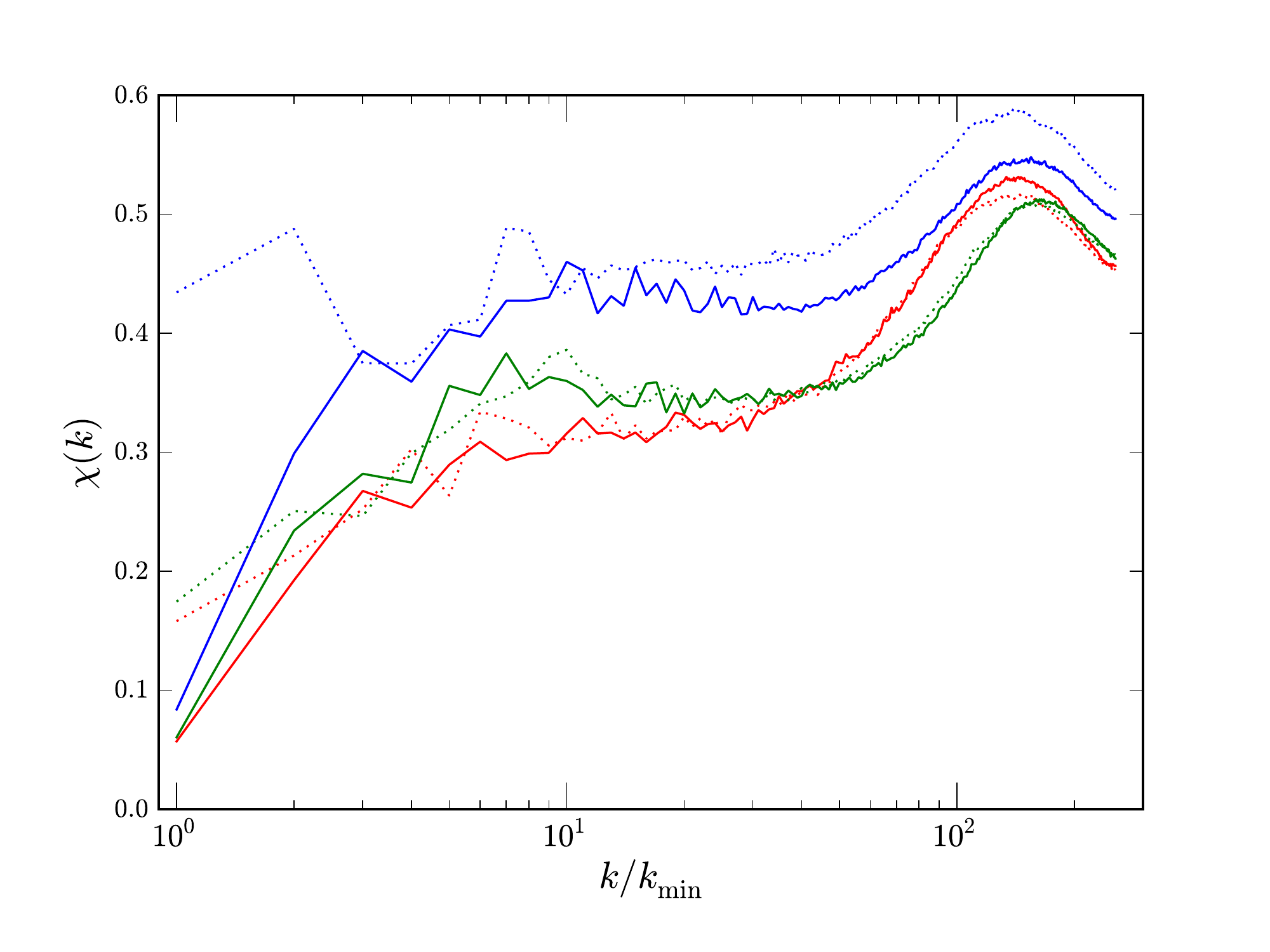}
\else
\includegraphics[width=\hw\textwidth]{figs_spectra//Helmholtzb025vb25vb205vconverging-velocity_10_63.eps}
\fi
\caption[ ]{The ratio of compressible to solenoidal motions,
$\chi(k)=P(\vc,k)/P(\vs,k)$, for all three simulations (red, green and blue for \betal,
\betam, \betah) and two snapshots ($t=0.1, 0.6 \tff$, solid and dotted lines,
respectively)}
\label{fig.HelmholtzBoth} \end{center} \end{figure}

The ratio of compressible to solenoidal power, $\chi(k)$,  is shown in Figure \ref{fig.HelmholtzBoth} for the same fiducial snapshots
as in Figure \ref{fig.VrhoBoth}.  For \betah, we find $\chi(k)\approx0.4$ in the
range $\kkmin\in[2,30]$, which is
less than $0.5$ as expected from purely geometrical considerations: for a given
wave vector $\bf{k}$ there is only one longitudinal (compressional) component,
while there are two transverse (rotational) components.  This is consistent
with the findings of \citet{Kritsuk10b}, where the increase of the magnetic
field strength decreases $\chi(k)$.  
Again we find the increased compressibility
with increased $\betao$.  At low $\kkmin<50$ there is a monotonic increase in
$\chi(k)$ with $\betao$.  The \betah\ simulation also shows a substantial low
$\kkmin<2$ compression at late times, seen to a lesser extent in the other two.  The $\betam$
run shows a peak in $\chi(k)$ at $\kkmin\approx 10$, probably with a similar
origin as the $\kkmin<2$ increases, namely a result of the large scale
forcing pattern.  
In all cases, there is a general trend of increasing compressibility at
higher $\kkmin$.  As discussed in Section \ref{sec.Pv}, this increase in
compressibility causes reduced amplification of the magnetic field at high
$\kkmin$.

\subsection{Helmholtz Decomposition in the Collapsing State}
There is no substantial increase in compressive
motions at high $\kkmin$ as time progresses, indicating, as for the velocity
power spectrum, $P(v,k)$, that this statistic is not especially sensitive to
gravity, due to its volume-weighted nature.
It seems that the lower levels of magnetization in the \betah\ simulation is enough
for that run to have some increased compressional motion at high $\kkmin$.  The
stronger field runs show almost no evolution. At low $\kkmin$, all three
simulations show an increase in power at $\kkmin<2$.  This is possibly due to
the effects of gravity causing large scale contraction of the gas.  

\section{Discussion}\label{sec.discussion}\label{sec.WhereIPutSFR}

The combined effects of gravity, turbulence, and magnetic fields have yet to be
incorporated in star formation theory with their proper respective weights.  
Here we discuss modifications to two recent star formation models (Section
\ref{sec.theory}) and interpret several recent observations in this light
(Section \ref{sec.observations}).

\subsection{Implications on Theory}\label{sec.theory}

In the past decade, the lognormal density PDF of turbulence has been used to
predict a number of properties of star formation, including the star
formation rate.  Here we will examine the implication on two works in particular,
namely the star formation rate models of \citet{Krumholz05}, hereafter KM05, and
\citet{Padoan11}, hereafter PN11.  In both of these works, the star formation rate per free fall
time, \sfrff, is predicted to be proportional to the mass fraction above some
critical density, using a lognormal distribution.  Thus
\begin{align}
\sfrff &= \epsilon \int_{\xcr}^\infty \rho V(\rho) d\rho \\
& = \epsilon \left (1+\rm{erf}\left[\frac{\sigma^2 - 2 \ln \xcr}{2^{3/2}\sigma} \right]
\right), \label{eqn.erf}
\end{align}  
where we have used Equation \ref{eqn.lognormal} in the derivation of the second
of these.  This model presumes that the effects of gravity do not alter the
structure of the star-forming cloud.  However if the effects of gravity are
present during the formation of the cloud, it is not unreasonable to suppose
that a power-law density PDF might be present before the onset of turbulence.
Here we will explore what effects the distribution may have on the star
formation rate.

KM05 and PN11 differ in two ways:  the nature of $\epsilon$ and the
nature of \xcr.  The parameter $\epsilon$ contains the timescale for collapse,
as well as the fraction of collapsing gas reaching the proto-star.  
The other difference is the
nature of \xcr.  In KM05, \xcr\ is the density at which velocity perturbations
become subsonic and lose their turbulent support.  In PN11, \xcr\ is the post
shock density that yields clumps that are larger than the sum of the
Bonnor-Ebert \hlthree{and magnetic critical masses.}

\def\lams{\ensuremath{\lambda_{\rm{s}}}}
\def\lamj{\ensuremath{\lambda_{\rm{J}}}}
As we have shown in Section \ref{sec.Pv}, the scaling of the turbulence 
depends on the mean magnetic field.  This in turn has consequences for the
expected critical density.  The KM05 result predicts that 
\begin{align}
    \hlthree{\xcr = \left(\phi \frac{\lamj}{\lams}\right)^2},\\
\lambda_s = L_0 (\cs/\sigma_0)^{1/q},
\end{align}
thus
\begin{align}
    \hlthree{\xcr = \left(\phi \frac{\lamj}{L_0}\right)^{2} \mach^{2/q}}
\end{align}
where $\lams$ is the sonic length, at which velocity perturbations become
subsonic, \hlthree{$L_0$ is the cloud size}, $\sigma_0$ is the velocity fluctuation at
the outer scale, and $\phi$ is a numerical factor of order unity.  In their work
they find $\phi=1.12$.  We can relate $q$ to the velocity spectral index $\pvex$ by
$q=-(\pvex+1)/2$.  For $\pvex= -2$, as expected for pressure free
turbulence, $q=1/2$, while for $\pvex=-1.5$ as seen in our $\betal$ run, $q=1/4$.
If we take $\mach=10$, we find an
increase of several orders of magnitude in $\xcr$ when using the spectral
scaling of the \betal\ simulation over the \betah\ simulation, assuming
\hlthree{everything else stays constant}.  In PN11, \xcr\ does not
depend directly on $\pvex$, but will likely impact the post-shock magnetic field
distribution, similarly increasing \xcr\ for increasing mean field strength.  

Another impact of our results here is the result of replacing the turbulent PDF, $V(\rho)$, with one
that transitions to a power-law at some transition density $\rho_t$.  Thus,
\def\vp{\ensuremath{V_{\rm{p}}}}
\begin{align}
\vp(\rho) d \ln\rho  = 
\begin{cases}
 N\frac{1}{\sqrt{2\pi}\rho\sigma}\rm{exp}\left[\frac{( \ln\rho - \mu)^2}{2
 \sigma^2} \right ] d\ln\rho, & \rho < \rhot \\
 N\rho_0 \rho^{\vrex} d\ln\rho, & \rho > \rhot ,
\end{cases}
\label{eqn.piecewise}
\end{align}
rather than a plain lognormal.  Here the normalization $N$ is determined by the
normalization requirement,
$\int_{-\infty}^{\infty} \vp(\rho) d\ln\rho=1$, and is given by
\begin{align}
  N=2 \left(1+\rm{erf}\left( \frac{2 \ln \rhot
  +\sigma^2}{2^{3/2} \sigma} \right) - \frac{2 \rho_0\rhot^{\vrex}}{\vrex}
  \right)^{-1}.
\end{align}
From here we are left to determine the critical density for star formation,
$\rhoc$, and the density at which the PDF transitions from lognormal to
power-law, $\rhot$. These are not necessarily equal; the forces of
rotation and magnetic fields will suppress the collapse of material once the gas
is within the potential well of the self-similar sphere, whereby $\rhoc$ may be
larger than $\rhot$.  
If we assume that
$V(\rho)$ is continuous and differentiable, we can piece together an analytic
estimate for $\rho_0$ and $\rhot.$  Differentiability is not necessarily the
correct condition to take, but in the absence of a better condition for the
transition density $\rhot$, an assumption must be made.  The PDF in Figure
\ref{fig.VrhoBoth} as well as some observations do appear to be differentiable.
Nevertheless, we can use this assumption to
examine the sensitivity of $\sfrff$ to $\rhot$ and $\rho_0$.   These two conditions give us
\begin{align}
\rhot&=\frac{1}{2}(2\vrex -1)\sigma^2\\
\rho_0&=e^{\sigma^2 \vrex (\vrex - 1)/2}.
\label{eqn.rho0}
\end{align}

Using the fit values from Table
\ref{table.lognormal_fits}, we find values listed in Table
\ref{table.transition_fits}.   We also compute values
for the Larson-Penston sphere, with $\rho\propto r^{-2}$ and $\sigma=1.5$ as
found for the majority of our simulations, and a pressure-free collapse with
$\sigma$ taken from Equation \ref{eqn.lognormalWidth} for $b=0.4$ and $\mach=9$.  

\begin{figure} \begin{center}
\ifpdf
\includegraphics[width=\hw\textwidth]{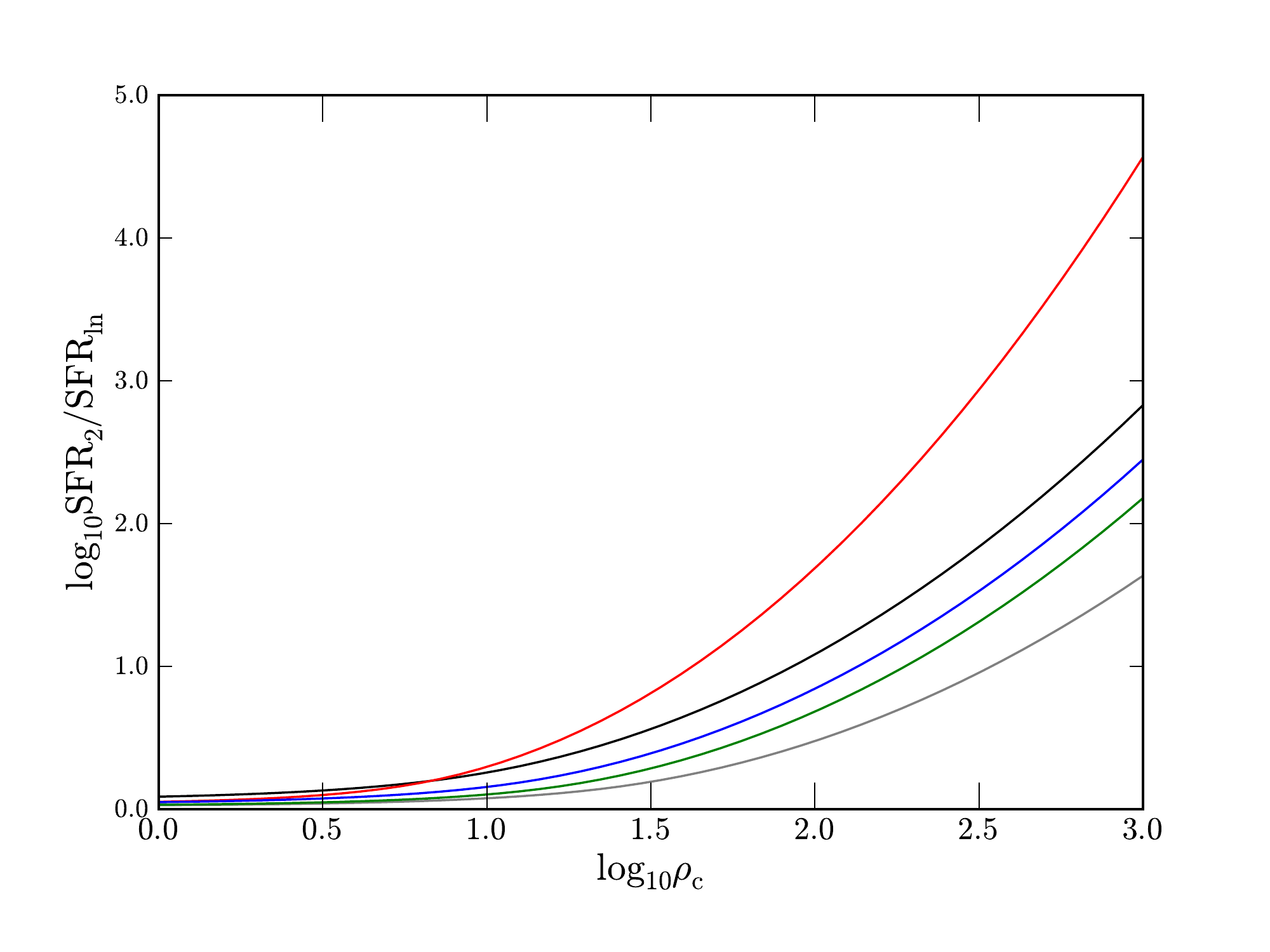}
\else
\includegraphics[width=\hw\textwidth]{figs_rate/SFRmod.eps}
\fi
\caption[ ]{Excess cumulative mass fraction of a power-law PDF relative to a
lognormal PDF.  This gives the amount by which the two primary star formation
rate models are low. The black line uses $\sigma=1.5,\vrex=-1.5,\rho_0=0.5$.
Red, green, and blue lines use $\sigma$ taken from Table
\ref{table.lognormal_fits}, $\vrex$ from Figure \ref{fig.vrex_convergence}, and
$\rho_0$ from Equation \ref{eqn.rho0}.  The grey line uses a pressure-free
solution (PF)}
\label{fig.SFR} \end{center} \end{figure}

The increase in predicted star formation rate is found by using this
piecewise PDF instead of a lognormal in Equation \ref{eqn.erf}.  
In Figure \ref{fig.SFR}, we plot the piecewise cumulative mass fraction in
Equation \ref{eqn.piecewise} relative to the purely lognormal mass fraction
versus critical density $\rhoc$.  We have used parameters from all three
simulations, red, green, and blue lines for \betarespectively, the
Larson-Penston solution in black, and the pressure free solution in grey.  
We find that the increase in
star formation rate can be quite large, and is quite sensitive to the details of
$\rhot$ and $\vrex$.  This clearly predicts the incorrect behavior in rate
relative to $\betao$, as the $\betal$ model has a measured collapse rate lower
than the other two (see Figure \ref{fig.sfr}) .  This is due to the selection of $\rhot$, which is
significantly lower for $\betal$ than the other two, caused by the lower value
of $\sigma$.  
While this is an incorrect prediction, it demonstrates how
sensitive this type of model is to the parameters of the fit.  In reality,
$\xcr$ and $\rhot$ will be functions of a stability criterion that fully
incorporates magnetic fields and turbulence.  

This result is consistent with the findings of \citet{Cho11}, who used $512^3$
unigrid simulations to examine the
collapse rate for various values of critical density.  For the largest critical
values, $\rhoc=500\rho_0$, they find an increase of 2400 relative to the
predicted value of KM05.  

Future work will analyze bound clumps to determine a proper value for \xcr\ and
\rhot, and further evaluate this prediction as well as KM05 and PN11.

\begin{center}
\ctable[pos=h,caption={Parameters for extended density
PDF},label={table.transition_fits}]{ccccc}{}{
\hline
\hline
model & $\vrex$ & $\sigma$ & $\rho_0$ & $\rhot$ \\
\hline \\
0.2 & 1.2 & 1.75 & 2.4 & 6\\
2   & 1.5 & 1.75 & 3.2 & 17\\
20  & 1.5 & 1.64 & 2.4 & 13 \\
PF  & 1.8 & 1.71 & 3.4 & 24\\
LP  & 1.5 & 1.5  & 1.9 & 10\\
\hline
}
\end{center}

One question that is not addressed by this work is when star formation begins
in the lifetime of a cloud.  That is, are the effects of self gravity already
present before the lognormal PDF is established, in which case a power-law
$V(\rho)$ is the best picture to take, or is a star-forming cloud turbulent
first, and then transitions to self-gravitating through cooling?
\hlthree{\citet{Kainulainen09} two families of clouds, one with only lognormal PDFs and no
star formation, and one family that includes power-law tails and high rates of star formation.
However it is presently unknown if the star formation begins before or after
the power-law tail, or if the tail impacts the star formation itself.}


\subsection{Interpretation of Observations}\label{sec.observations}

The role of magnetic fields has been vigorously debated over the last few
decades.  Initially magnetic fields were the dominant mechanism for regulating
star formation
\citep{Mouschovias76,Shu87}, then relegated to an insignificant role as
supersonic turbulence took hold \citep{MacLow04,Krumholz05}.  Recent
observations have given a mixed review of the role of magnetic fields, with some
measurements indicating a strong field, and some indicating a weak field.  Here
we will discuss the implications of our present model on recent observations.

\citet{Li09} measured starlight polarization in a number of molecular clouds, and averaged said
polarization over large (30 pc) scales of the cloud and small (0.3 pc) scales
centered on high density cloud cores. \hlthree{The resultant alignment of magnetic fields
on large and small scales is
interpreted to imply \suba\  magnetic fields, by way of comparison to a set of
\sa\ ($\alfmach = 2$) and \suba\ ($\alfmach = 0.7$) simulations.  In sub-\Alfvenic
turbulence, the kinetic energy of the gas is unable to alter the alignment of
the magnetic field at all scales, leading to alignment with the mean field at
all scales and a correlation between direction at both large and small scales.
In contrast, \sa\ turbulence  can greatly alter the magnetic field direction,
reducing the correlation between large and small scales. This is
consistent with the increase in mean alignment between
$\vvec$  and $\bvec$ with decreasing $\betao$ (Figure \ref{fig.BVangle}), as in
our \transa\ simulation there is a trend towards alignment that is absent
from the higher $\betao$ runs.  In our \transa\ simulation, kinetic energy
is insufficnent to alter the direction of the field relative to the mean, and in
turn velocity and magnetic vectors are aligned.  In the \sa\ simulations, on the
other hand, kinetic motions dominate and the field direction can be altered by
the velocity.  On the other hand, recent measurements of the directions of
outflow from T-Tauri}
stars are uncorrelated with mean magnetic field \citep{Menard04,Targon11},
though there is a weak correlation for earlier Class 0 and Class 1 objects
\citep{Targon11}.  
Finally, two sets of observational papers show a transition in field strength
with scale.  \citet{Crutcher10}  find that density is uncorrelated with field
strength for densities  $n_{\rm{H}} < 10^3$, and correlated as $B\propto n^{0.65}$
for higher densities, using Zeeman splitting.   \citet{Heyer12} use velocity
anisotropy in the Taurus molecular cloud to show that the low column density
gas, with more anisotropic flows, is likely \transa, while high column
density gas, with more isotropic flow, is likely \sa.  

This suite of measurements is consistent with what we have seen in this study,
and can be understood by the combination of Figures \ref{fig.Pdyn},
\ref{fig.BVangle}, and \ref{fig.PengBoth},
  focusing on the $\betal$ simulation.  Through the action of supersonic
turbulence, the effect of magnetic fields is a function of scale.  Low $\kkmin$
gas, where the turbulence is generated in the presence of an appreciable mean
field, comes into equipartition.  This allows low density gas to exhibit
alignment seen in observations.  However in material at higher $\kkmin$, where the
structures are generated by shocks, the field is unable to come
into equipartition.  This high density gas is further selected to be \sa\ in
post-shock gas where the magnetic energy is too weak to resist compression.

\begin{figure*} \begin{center}
\ifpdf
\includegraphics[width=\hw\textwidth]{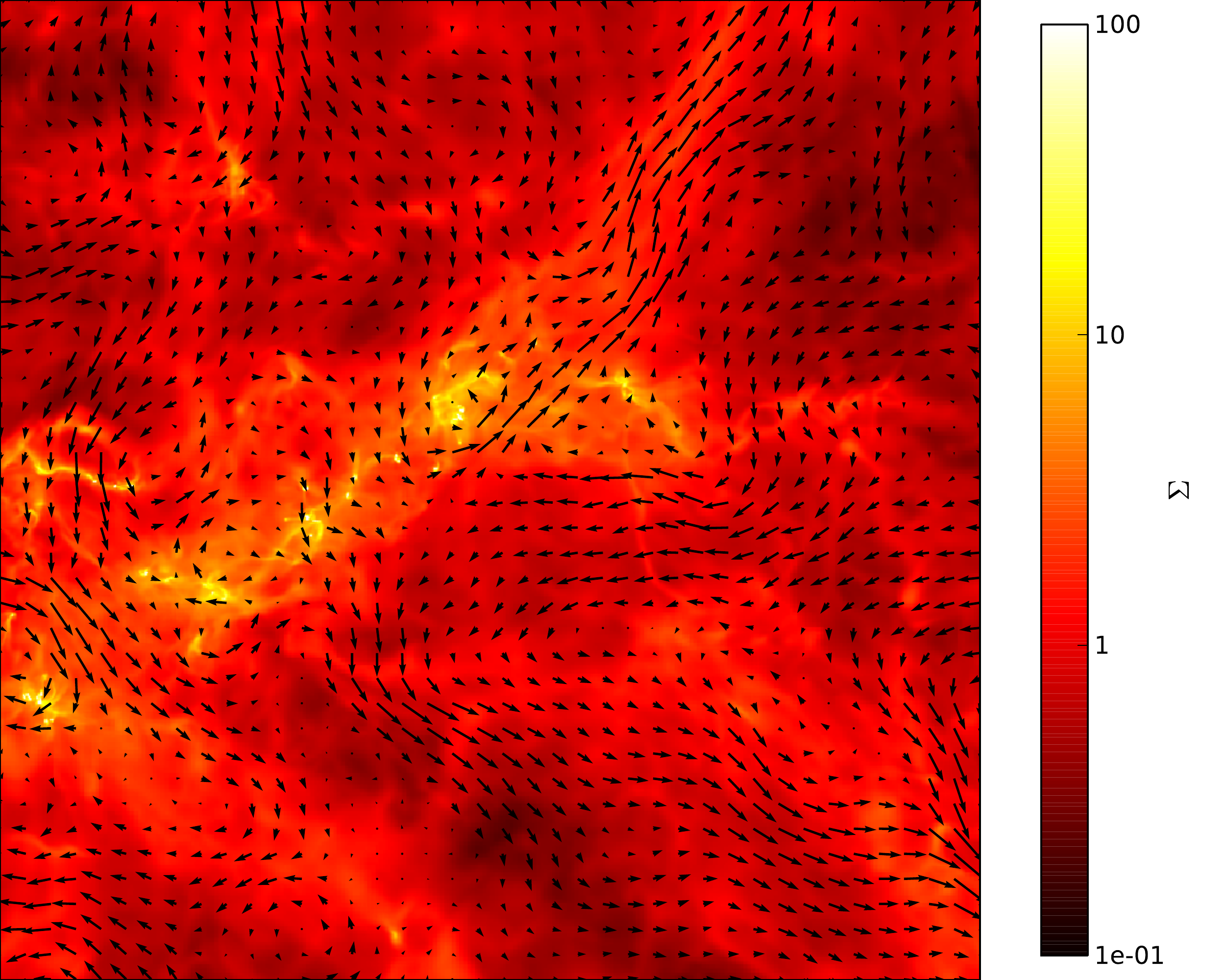}
\includegraphics[width=\hw\textwidth]{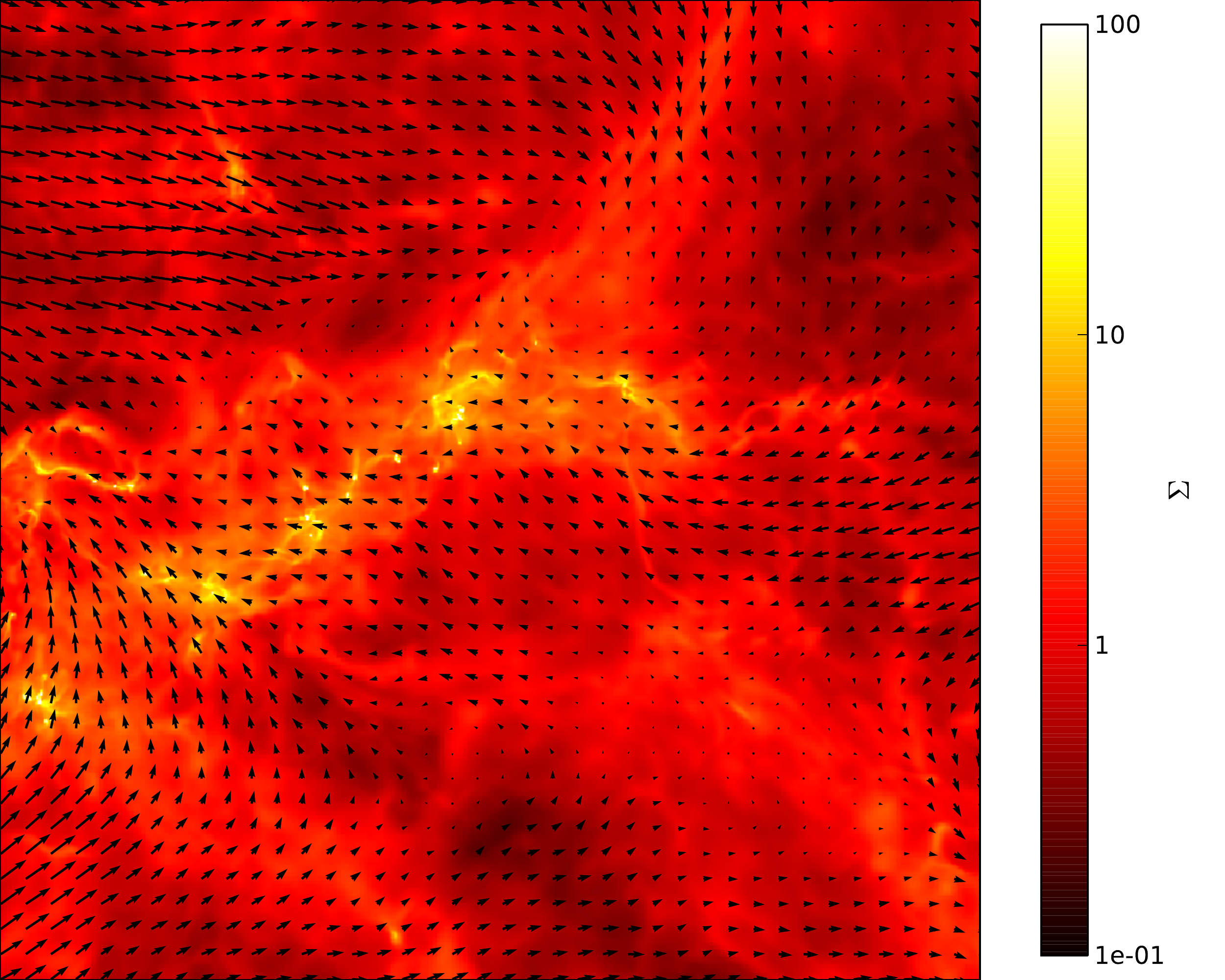}
\else
\includegraphics[width=\hw\textwidth]{figs_images//b25v_0061_xproj_magvec_Projection_x_Density.eps}
\includegraphics[width=\hw\textwidth]{figs_images//b25v_0061_xproj_velvec_Projection_x_Density.eps}
\fi
\caption[ ]{Projections of a sub set of dense gas from the $\betam$ simulation
at $\tlate$.  Overlaid are magnetic vectors (left) and velocity vectors (right).
The dense parts of this filament lie at the convergence of opposing velocity
streams, where the magnetic field is roughly aligned to the velocity.  Higher in
the stream, where the field is perpendicular to the velocity, collapse is
suppressed.}
\label{fig.filament} \end{center} \end{figure*}

We also find that our simulations are filled with filamentary structures, as seen by
\citet{Men'shchikov10} with \emph{Hershel} in the Aquila and Polaris clouds, see
Figures \ref{fig.projections} and \ref{fig.filament}.  Figure \ref{fig.filament}
shows a restricted region of the $\tlate$ snapshot of the $\betam$ simulation,
overlaid with magnetic field vectors (right panel) and velocity vectors (left
panel).
High density material, with $\rho>10$, is nearly all concentrated in
filamentary structures, as seen in \citet{Arzoumanian11}. Further analysis is necessary to quantify the filament width
distribution, as in that work filaments cluster tightly around $0.1 \rm{pc}$.
Magnetic fields are weak and perpendicular to the structure with the cores,
while in the lower density filamentary tail the field is parallel to the
structure.  This is expected by the fact that all high density material is \sa:
the filament aligned with the magnetic field will have maximally amplified the
magnetic field, halting the collapse, while the material with $\bvec$ and
$\vvec$ more closely aligned has lower supporting pressure due to the reduced
field amplification, and can go on to form ``stars''.  

We can make further contact between the collapsing state and observations of
prestellar cores seen in the Aquila star-forming cloud, as reported by
\citet{Andre10}.  We can estimate the mass-size relation from a self-similar
sphere as 
\begin{align}
  M( R )=4\pi\int_0^R\rho( r )r^2dr\propto R^{n+3}
\end{align}
which gives us $M(r)\propto r^{1}$ for the LP solution, $M(r)\propto r^{1.3}$
for the PF solution \hlthree{(see also \citet{Kritsuk11d})}.  The value found in
Aquila is 1.13 (Philippe Andr\' e, private communication.), which is somewhat
shallower than that of the LP solution.
For completeness, the expansion wave solution predicts 
$M(r)\propto r^{1.5}$, though is less relevant due to the fact that the
expansion wave solution presupposes a central singularity, while the Andr\' e
result focused on starless cores.

\section{Conclusions}\label{sec.Conclusion}


In this work, we use high resolution AMR MHD simulations to examine the combined
effects of self-gravity and magnetic fields on supersonic turbulence.  We find
that self-gravity bifurcates the cloud into two phases:  a low density,
turbulent cloud; and high density, self-gravitating cores.  The two phases have
substantially different statistical properties.  The magnetic field serves to
effectively decrease compressibility of the gas as the mean magnetic field is
increased.

The simulations were performed using the AMR Enzo code \citep{Bryan95,O'Shea04},
extended to MHD by \citet{Collins10}.  The three simulations use an rms Mach
number $\mach= 9$, virial parameter $\alpha=1$, and three values of the mean
magnetic field, giving initial plasma beta $\betao=0.2,2,20$.  The coarse level
had $512^3$ zones, and 4 levels of refinement were used, refining to keep the
Jeans length resolved by 16 zones.  This gives an effective resolution of
$8196^3$.  

The low density turbulent state exhibits \hlthree{properties anticipated by earlier
works in supersonic hydro and MHD turbulence}
The effects of increasing mean magnetic field strength
are to decrease \hlthree{both the} compressibility and the ability of the gas flow to bend field
lines.  The properties of the turbulent state can be summarized as follows:
\begin{itemize}
\item Density PDF, $V(\rho)$, can be roughly represented by a lognormal, given by
Equation \ref{eqn.lognormal}.  The width of the lognormal increases somewhat
with increasing mean thermal-to-magnetic pressure ratio, $\betao$.
\item The density power spectrum, $P(\rho,k)\propto k^\prex$, tends to flatten
with increasing Mach number.  We find slopes
that are flatter than other measurements of $\prex$ in supersonic turbulent
simulations, in proportion to the higher Mach number relative to earlier work.
We find that $\prex$  weakly decreases with increasing
$\betao$. For the early snapshot, $\prex=-0.42, -0.58,$ and $-0.62$
for \betarespectively;  
\item The column density power spectrum, $P(\Sigma,k)\propto
k^\psigmaex$, is also flatter than observed. We find $\psigmaex=-1.39, -1.61,$ and
$-1.53$ for \betarespectively, while values of $\psigmaex<-2.25$ have been
reported elsewhere.  This discrepancy is likely due to the decreased density
range in earlier measurements, in part due to the reduced dynamic range
available to the single tracer molecules used for the observations.
\item Thermal-to-magnetic pressure ratio, $\betath$, shows a decreasing
correlation between $\betath$ and $\rho$ with increasing $\betao$.  The average
slope also decreases with decreasing field. For $\betath\propto\rho^\betarex$,
$\betarex = 0.97,\ 0.77,$ and $0.54$ for \betarespectively.
\item Dynamic-to-magnetic pressure ratio, $\betadyn$, shows only mild
decrease in correlation between $\betadyn$ and $\rho$, and the peak of the
distribution decreases with increasing $\betao$.
\item The magnetic field and velocity are nearly aligned for mid- to low-density
gas in the strong field case, and decorrelated for the other two cases.
\item the PDF of magnetic field, $V(b)\propto \exp[-(b/b_0)^c]$, shows a stretched exponential with a
stretching exponent of $c\approx 1/3$, as given in Equation
\ref{eqn.stretched}, for the high field strength gas. The slope decreases with \betao, with $b_0=0.15, 0.19$ and
$0.23$ for \betarespectively, and the peak decreases with
increasing \betao.
\item The velocity power spectra, $P(v,k)\propto k^\pvex$, show an increasing
slope as \betao\ increases, with the weak field case resembling supersonic
hydrodynamic turbulence; we find that $\pvex = -1.46,\ -1.58$, and $-1.80$ for
\betarespectively.
\item Equipartition between kinetic and magnetic energy is only seen at
large scale for the strongest field case, $\betal$, and a small range of
intermediate wavenumbers for $\betam$;  the magnetic energy is an order of
magnitude or more for all high wavenumbers in all simulations, and all
wavenumbers in the weak-field case, $\betah$.
\item The ratio of compressive-to-solenoidal motions, $\chi(k)$, increases with
$\betao$, indicating increased compressibility with decreased magnetic field
strength.
\end{itemize}

The high-density collapse phase exhibits properties consistent with spherically
symmetric isothermal collapse solutions,
 with power-law density profiles $\rho\propto r^\rrex$.  Again the
effect of the magnetic field is to decrease the compressibility of the gas.  The
properties of the collapsing state can be summarized as follows:
\begin{itemize}
\item The density PDF is well approximated by a power-law, $V(\rho)\propto
\rho^\vrex$, where $\vrex=3/\rrex$, and $\vrex=-1.80,\ -1.78,$ and $-1.65$ for
\betarespectively.   The mass flux rate, which can be seen as a proxy for star
formation rate, is an increasing function of $\betao$.  The slope in the strong
field case is continuing to increase at the end of the simulation, due to the slower
collapse in this case.
\item The density power spectrum, $P(\rho,k)\propto k^\prex$, shows a positive
  slope,
 consistent with the expected behavior of a self-similar solution,
$\prex=-2(\vrex+1)$, though detailed comparison to theory is difficult due to
the superposition of the turbulent state. We find that $\prex=0.86, 1.12$, and
$1.2$ for \betarespectively.
\item The column density spectrum $P(\Sigma,k)\propto k^\psigmaex$, shows flat
spectra, with $\psigmaex\approx 0$ for all simulations;  this behavior is
potentially observable, provided observational tracers with high enough dynamic
range are used.
\item The thermal-to-magnetic pressure ratio, $\betath$, is near unity for all
values of $\betao$.
\item The dynamic-to-magnetic pressure ratio, $\betadyn$, is larger than unity
by at least two orders of magnitude for all values of $\betadyn$.
\item The high density gas shows a decrease in the mean alignment between
$\bvec$ and $\vvec$ with increasing density.
\item The magnetic field PDF shows a power-law behavior, $V(B)\propto B^\vbex$, with
$\vbex=-5.42, -4,$ and $-3.22$ for \betarespectively.
\item The velocity power spectra show no evidence of the collapsing state, due
  to the small volume fraction of the collapsing gas and lack of severe increase
  in velocity, as seen in the density power spectrum.
\item The kinetic energy spectra, $P(\rho^{1/2} u,k)$, and $P(B,k)$ show
increases in power at high $\kkmin$, consistent with the density weighted nature
of these two statistics.
\item The ratio of compressive-to-solenoidal power similarly shows almost no
signature of the collapsing state.
\end{itemize}

We find that for certain values of the transition from lognormal to power-law
density PDF, the feedback on the predicted star formation rate can be quite
large.  This has sensitive dependence on the transition density, $\rhot$ between
lognormal to power-law, and the critical density, $\rhoc$, above which cores are
considered to be gravitationally bound.  Additionally the flattening of the
velocity power spectrum in the \transa\ simulation will cause high density
material to remain gravitationally bound to smaller scales, increasing $\rhoc$
from values expected from hydrodynamic scaling.

High density collapsing gas is created very \sa, with $\alfmach\approx 100$, even in the
\transa\ simulation.  In the \transa\ simulation, we find large scale imprints
of the magnetic field, and at small scale this imprint weakens.  This is
consistent with a number of recent observations of the properties of
star-forming clouds.  High density cores seem to be found primarily in
filaments, in regions where the magnetic field is aligned perpendicular to the
filament.  Filaments with longitudinal magnetic fields do not show high density
material, consistent with the \sa\ nature of the collapsing state being a
necessary condition for collapse.

This work was supported in part by the National Science Foundation under grants
AST0808184 and AST0908740;  D.C., H.L. and H.X. were supported in part by Los Alamos
National Laboratory, LLC for the National Nuclear Security Administration of
the U.S. Department of Energy under contract DE-AC52-06NA25396. 
A.K. was supported in part by the National Science Foundation grant AST-1109570.
H.L. gratefully acknowledges the support of the U.S. Department of Energy
through the LANL/LDRD program.  Computer time was provided through  NSF TRAC
allocations TG-AST090110 and TG-MCA07S014.  The computations were performed on
Nautilus and Kraken at the National Institute for Computational Sciences
(http://www.nics.tennessee.edu/).




\bibliographystyle{apj}
\bibliography{apj-jour,ms.bib}  
\end{document}